\documentclass[twocolumn,showkeys,showpacs,preprintnumbers,prd,superscriptaddress,nofootinbib]{revtex4-1}
\usepackage{verbatim}
\usepackage[T1]{fontenc}
\usepackage[utf8]{inputenc}
\usepackage[american]{babel}
\usepackage{epsfig}
\usepackage{booktabs}
\usepackage{multirow}
\usepackage{dcolumn}
\usepackage{amsmath}
\usepackage{mathtools}
\usepackage{amsfonts}
\usepackage{amssymb}
\usepackage{ulem}
\usepackage{epstopdf}
\usepackage{bm}
\usepackage{siunitx}
\usepackage{braket}
\usepackage{enumitem}
\usepackage{soul}
\usepackage[table]{xcolor}
\usepackage{color}
\usepackage{transparent}
\usepackage{pifont}
\usepackage{enumitem}

\definecolor{navyblue}{rgb}{0.0, 0.0, 0.5}
\definecolor{royalblue}{rgb}{0.25, 0.41, 0.88}
\definecolor{cadmiumgreen}{rgb}{0.0, 0.42, 0.24}
\definecolor{blue-violet}{rgb}{0.54, 0.17, 0.89}
\definecolor{darkviolet}{rgb}{0.58, 0.0, 0.83}
\definecolor{orange(colorwheel)}{rgb}{1.0, 0.5, 0.0}

\usepackage{hyperref}
\hypersetup{
    colorlinks=true, 
    linkcolor=royalblue, 
    citecolor=magenta}

\usepackage{booktabs}
\usepackage{multirow}
\usepackage{dcolumn}
\usepackage{colortbl}

\makeatletter\let\expandableinput\@@input\makeatother

\begin{document}
\title{Mapping the $\Lambda_{\rm s}$CDM scenario to $f(T)$ modified gravity:\\ Effects on structure growth rate}

\author{Mateus S. Souza}
\email{mateusscherersouza@gmail.com}
\affiliation{Instituto de F\'{i}sica, Universidade Federal do Rio Grande do Sul, 91501-970 Porto Alegre RS, Brazil}

\author{Ana M. Barcelos}
\email{acmb.ufrgs@gmail.com}
\affiliation{Instituto de F\'{i}sica, Universidade Federal do Rio Grande do Sul, 91501-970 Porto Alegre RS, Brazil}

\author{Rafael C. Nunes}
\email{rafadcnunes@gmail.com}
\affiliation{Instituto de F\'{i}sica, Universidade Federal do Rio Grande do Sul, 91501-970 Porto Alegre RS, Brazil}
\affiliation{Divis\~ao de Astrof\'isica, Instituto Nacional de Pesquisas Espaciais, Avenida dos Astronautas 1758, S\~ao Jos\'e dos Campos, 12227-010, SP, Brazil}

\author{\"{O}zg\"{u}r Akarsu}
\email{akarsuo@itu.edu.tr}
\affiliation{Department of Physics, Istanbul Technical University, Maslak 34469 Istanbul, T\"{u}rkiye}

\author{Suresh Kumar}
\email{suresh.kumar@plaksha.edu.in}
\affiliation{Data Science Institute, Plaksha University, Mohali, Punjab-140306, India}

\begin{abstract}
The concept of a rapidly sign-switching cosmological constant, interpreted as a mirror AdS-dS transition in the late universe and known as the $\Lambda_{\rm s}$CDM, has significantly improved the fit to observational data, offering a promising framework for alleviating major cosmological tensions such as the $H_0$ and $S_8$ tensions. However, when considered within general relativity, this scenario does not predict any effects on the evolution of the matter density contrast beyond modifications to the background functions. In this work, we propose a new gravitational model in which the background dynamics predicted by the $\Lambda_{\rm s}$CDM framework are mapped into $f(T)$ gravity, dubbed $f(T)-\Lambda_{\rm s}$CDM, rendering the models indistinguishable at the background level. However, in this new scenario, the sign-switching cosmological constant dynamics modify the evolution of linear matter perturbations through an effective gravitational constant, $G_{\rm eff}$. We investigate the evolution of the growth rate and derive new observational constraints for this scenario using RSD measurements. We also present new constraints in the standard $\Lambda_{\rm s}$CDM case, incorporating the latest Type Ia supernovae data samples available in the literature, along with BAO data from DESI. Our findings indicate that the new corrections expected at the linear perturbative level, as revealed through RSD samples, can provide significant evidence in favor of this new scenario. Additionally, this model may be an excellent candidate for resolving the current $S_8$ tension.

\end{abstract}

\maketitle
\section{Introduction}
\label{sec:intro}

Several extensions of general relativity (GR) have been proposed and extensively studied to address key observational challenges in cosmology and astrophysics (see \cite{Clifton_2012, Ishak_2018, Nojiri_2017, saridakis2023modifiedgravitycosmologyupdate, Frusciante_2020} for reviews). Notably, modified gravity (MG) models, which introduce additional gravitational degrees of freedom, extend the standard $\Lambda$CDM framework and can account for the accelerated expansion of the Universe at late times. Although many of these models fit observational data well, they often lead to theoretical degeneracies where different models yield similar observational signatures, complicating their differentiation. Among the viable MG theories, those based on torsion, particularly the Teleparallel Equivalent of General Relativity (TEGR) \cite{Maluf_2013}, have gained significant attention due to their unique formulation of gravity using torsion instead of curvature. In TEGR, where the Lagrangian is represented by the torsion scalar $T$, the simplest extension is $f(T)$ gravity, which generalizes the Lagrangian to a non-linear function of $T$ (see \cite{Bahamonde_2023, Cai_2016, Kr_k_2019} for reviews). This extension introduces additional degrees of freedom that can potentially address current cosmological tensions, such as the discrepancies in the Hubble constant $H_0$ and the amplitude of matter fluctuations $S_8$, making $f(T)$ gravity an attractive candidate for probing deviations from GR.

Increasingly precise measurements of cosmological parameters are challenging the consensus on the $\Lambda$CDM model \cite{Abdalla_2022}. The most prominent discrepancy concerns the current rate of cosmic expansion, quantified by the Hubble constant, $H_0$. Analysis of Planck-CMB data within the minimal $\Lambda$CDM framework yields $H_0 = 67.4 \pm 0.5$ km s$^{-1}$ Mpc$^{-1}$ \cite{Planck:2018vyg}, which is in approximately $5\sigma$ tension with the local measurement reported by the SH0ES team, $H_0 = 73.30 \pm 1.04$ km s$^{-1}$ Mpc$^{-1}$ \cite{Riess:2021jrx}. Furthermore, multiple independent late-time observations also suggest higher values for the Hubble constant, reinforcing the significance of this tension (see discussions in \cite{Abdalla_2022, Di_Valentino_2021, Perivolaropoulos_2022}).

Cosmic shear surveys and Planck-CMB anisotropy measurements reveal another significant discrepancy related to the weighted amplitude of matter fluctuations, defined as $ S_8 = \sigma_8 \sqrt{\Omega_{\rm m} / 0.3} $, where $\sigma_8$ characterizes the amplitude of matter fluctuations on scales of 8 $ h^{-1} $ Mpc, $\Omega_{\rm m}$ is the present-day matter density parameter, and $ h $ is the dimensionless reduced Hubble parameter, defined as $ h = H_0 / 100 $ km s$^{-1}$ Mpc$^{-1}$. This so-called $ S_8 $ tension, which has direct implications for the growth of cosmic structures, has emerged alongside the well-documented $ H_0 $ tension, prompting significant interest in potential extensions to the standard $\Lambda$CDM paradigm (see reviews in \cite{Abdalla_2022, Di_Valentino_2021, Perivolaropoulos_2022}). The Planck-2018 CMB anisotropy analysis under the $\Lambda$CDM framework reports a best-fit value of $ S_8 = 0.834 \pm 0.016 $ \cite{Planck:2018vyg}, indicating a 2–3$\sigma$ discrepancy when compared to results from cosmic shear surveys \cite{Dalal_2023, KiDS:2020suj, Amon_2022}. Redshift Space Distortion (RSD) data further reveal a 2.2$\sigma$ tension with Planck-2018 findings, measuring $ S_8 = 0.762^{+0.030}_{-0.025} $ \cite{Nunes_2021}. When data from cosmic shear, real-space clustering, and RSD analyses are combined within a modified gravity context, the growth tension escalates, increasing from 3.5$\sigma$ (using only $ f\sigma_8 $ data) to 6$\sigma$ when $ E_g $ measurements are incorporated \cite{Skara_2020}. These persistent discrepancies, unlikely to be resolved by systematic errors alone, have motivated comprehensive investigations into whether new physics beyond the standard model could address these fundamental tensions.

On the other hand, it is well established that modifications to GR can significantly influence the growth of density fluctuations, the formation of large-scale structures, and CMB anisotropies, among other cosmological phenomena (see \cite{Clifton_2012, Ishak_2018, Frusciante_2020} for comprehensive reviews). The $f(T)$ gravity model, a prominent extension within modified gravity theories, has been extensively studied using geometric data to compute the modified expansion rate of the Universe, characterized by the Hubble parameter $H(z)$ \cite{Briffa_2022, Briffa_2023, sandovalorozco2024testingftcosmologieshii, Zhadyranova_2024, Capozziello_2017, Qi_2017, Basilakos_2018, El_Zant_2019, Said_2020, Benetti_2020, dos_Santos_2022, Aljaf_2022, Sabiee_2022, dos_Santos_2023, Kavya_2024, Nunes_2016}. Additionally, its implications on sub-horizon scales have been investigated through analyses of the growth rate of cosmic structures \cite{Capozziello_2024, Zhadyranova_2024, aguilar2024nonfluidlikeboltzmanncode, Briffa_2024, Anagnostopoulos_2019, Sandoval_Orozco_2024}, while full CMB datasets have been employed to explore its broader impacts on cosmic microwave background anisotropies \cite{Nunes_2018, Kumar_2023, Wang_2020}.

Beyond their ability to explain the late-time accelerated expansion of the Universe~\cite{Bengochea:2008gz, Linder:2010py}, $f(T)$ gravity models can lead to an effective dark energy (DE) that exhibits phantom and phantom divide line (PDL) crossing behaviors~\cite{Wu:2010av, Karami:2010bys, Bamba:2010wb, Cardone:2012xq}. The phenomenological DMS20 model~\cite{DiValentino:2020naf} proposed that a PDL crossing at $z \sim 0.1$ is a promising candidate for alleviating the $H_0$ tension. However, a recent examination~\cite{Adil:2023exv} showed that the DMS20 model's capacity to reach negative energy densities for $z \gtrsim 2$ and mimic a negative cosmological constant at higher redshifts plays a critical role in mitigating this tension. This aligns with the findings of the $\Lambda_{\rm s}$CDM model~\cite{Akarsu:2021fol, Akarsu:2022typ, Akarsu:2023mfb}, which suggests an Anti-de Sitter (AdS) to de Sitter (dS) transition in DE (interpreted either as an effective field arising from modified gravity or as an actual field within the framework of GR) at redshift $z_\dagger \sim 2$, as conjectured through the graduated dark energy (gDE) model~\cite{Akarsu:2019hmw}. It is important to highlight that the values of $z_\dagger \sim 2$ are not arbitrarily fixed. Estimates for the transition redshift are obtained by allowing $z$ to be a free parameter in robust statistical analyses using cosmological data (see \cite{Akarsu:2021fol, Akarsu:2022typ, Akarsu:2023mfb,Akarsu:2024qsi, Akarsu:2024eoo}). The $\Lambda_{\rm s}$CDM model has shown promise in addressing major tensions, such as those involving $H_0$ and $S_8$. Recently, the Dark Energy Spectroscopic Instrument (DESI) BAO data have provided evidence for dynamical DE with more than $2\sigma$ confidence when using the Chevallier–Polarski–Linder (CPL) parameterization~\cite{DESI:2024mwx}. Moreover, non-parametric DE reconstructions using the same data suggest that the DE density may become negligible or even negative for $z \gtrsim 1.5\text{--}2$~\cite{DESI:2024aqx, Escamilla:2024ahl}. This trend aligns with pre-DESI findings, including those derived from SDSS BAO measurements~\cite{Escamilla:2023shf, Sabogal:2024qxs, Escamilla:2024ahl}. The $\Lambda_{\rm s}$CDM model provides one of the most economical frameworks for such a scenario, having only one additional parameter compared to the standard $\Lambda$CDM model—the redshift of the AdS-to-dS transition $z_\dagger$. Despite theoretical challenges initially anticipated in realizing the $\Lambda_{\rm s}$CDM scenario, recent studies have proposed mechanisms to account for it, such as Casimir forces in dark dimension models~\cite{Anchordoqui:2023woo, Anchordoqui:2024gfa, Anchordoqui:2024dqc} and successfully embedding the $\Lambda_{\rm s}$CDM into a type II minimally modified gravity known as VCDM~\cite{Akarsu:2024qsi, Akarsu:2024eoo}. This embedding elevates the $\Lambda_{\rm s}$CDM model to a fully predictive framework. When considered within the framework of GR, the $\Lambda_{\rm s}$CDM model modifies the background dynamics relative to $\Lambda$CDM while preserving the equations of motion for perturbations. In contrast, the $\Lambda_{\rm s}$VCDM model, equipped with a well-defined Lagrangian, introduces modifications in both the background evolution and perturbative equations. Therefore, implementing the $\Lambda_{\rm s}$CDM framework in different theories, particularly when a smooth transition is considered, would differ at the level of linear perturbations even for the same background dynamics. Since observables depend on both the background evolution and cosmological linear perturbation dynamics, it is expected that different realizations of the $\Lambda_{\rm s}$CDM model will yield different constraints from observational data. This allows us to further study, distinguish, and choose among the theories in which the $\Lambda_{\rm s}$CDM scenario can be realized. Recently, it was shown through the exponential infrared $f(T)$ gravity model\cite{Awad:2017yod}, which shows considerable potential in addressing the cosmological $H_0$ tension~\cite{Hashim:2020sez,Hashim:2021pkq}, that there could be previously overlooked solution spaces holding even greater promise~\cite{Adil:2023exv}. Specifically, by relaxing the customary assumption of a strictly positive effective DE density—natural in general relativity—new possibilities arise. It was demonstrated that, ensuring consistency with CMB data, the model yields the widely studied case of phantom behavior, while the previously overlooked case features a sign-changing DE density that transitions smoothly from negative to positive values at redshift $z_\dagger \sim 1.5$, aligning with recent approaches to alleviating cosmological tensions. Following all these developments, it is compelling to attempt embedding the $\Lambda_{\rm s}$CDM scenario into teleparallel $f(T)$ gravity and investigate its feasibility. If possible, studying this embedding could be valuable, as even for the same background dynamics, it could introduce modifications in linear perturbations.

Building on these developments, the novel aspect of the present work is to map the background dynamics predicted by the $\Lambda_{\rm s}$CDM model into the framework of $f(T)$ gravity theories. While both models are equivalent at the background level, they differ in their predictions for linear perturbations. Given that the $\Lambda_{\rm s}$CDM class of models provides a better fit to observational data than the standard $\Lambda$CDM model and effectively addresses the $H_0$ tension, our aim is to construct an $f(T)$ gravity model whose background dynamics replicate those of the $\Lambda_{\rm s}$CDM model. This approach ensures that $f(T)$ gravity can also tackle the $H_0$ tension. However, this new scenario modifies the linear perturbations of matter beyond the alterations in the Hubble parameter $H(z)$, which are not accounted for in the standard $\Lambda_{\rm s}$CDM dynamics. In this article, we will quantify the growth rate of structures within this new model and derive new observational constraints using robust Redshift Space Distortion (RSD) datasets. A more comprehensive and detailed analysis of the proposed model will be presented in a future communication.

The manuscript is organized as follows. In Section \ref{model}, we briefly review the fundamental aspects of the $\Lambda_{\rm s}$CDM model and $f(T)$ gravity. We introduce a new parameter within this gravitational framework that governs the linear perturbations and present the $f(T)$-$\Lambda_{\rm s}$CDM model. In Section \ref{data}, we define the datasets and the statistical methodology employed to analyze the data used in this work. In Section \ref{results}, we present and discuss our main results, providing insights into the implications of the model. Finally, in Section \ref{final}, we conclude with a summary of the key findings and outline future perspectives for further investigation.


\section{The $\Lambda_{\rm s}$CDM scenario in a $f(T)$ gravity}
\label{model}

The $\Lambda_{\rm s}$CDM paradigm is inspired by a recent conjecture proposing that the universe underwent a spontaneous mirror AdS-dS transition, characterized by a sign-switching cosmological constant ($\Lambda_{\rm s}$) around $z \sim 2$~\cite{Akarsu:2019hmw, Akarsu:2021fol, Akarsu:2022typ, Akarsu:2023mfb, Yadav:2024duq}. This conjecture arose from studies of the \textit{graduated dark energy} (gDE) model~\cite{Akarsu:2019hmw}, which showed that a rapid, smooth transition from an AdS-like to a dS-like dark energy component at $z \sim 2$ could mitigate major cosmological tensions, such as the $H_0$ and BAO Ly-$\alpha$ discrepancies~\cite{Akarsu:2019hmw}. The $\Lambda_{\rm s}$CDM model modifies the standard $\Lambda$CDM by replacing the constant cosmological term ($\Lambda$) with a sign-switching cosmological constant, which can be represented by sigmoid-like functions, such as ${\rm sgn}\,x \approx \tanh{kx}$ for $k > 1$, with $x$ as the redshift ($z$) or scale factor ($a=1/(1+z)$) in a Robertson-Walker metric. A specific example is $\Lambda_{\rm s}(z) = \Lambda_{\rm dS} \tanh[\eta(z_{\dagger}-z)]$, where $\eta > 1$ controls the rapidity of the transition, and $\Lambda_{\rm dS} = \Lambda_{\rm s0}/\tanh[\eta\,z_\dagger]$. For transitions with $\eta \gtrsim 10$ around $z_\dagger \sim 1.8$, $\Lambda_{\rm dS} \approx \Lambda_{\rm s0}$ holds. In the limit $\eta \rightarrow \infty$, the model becomes the \textit{abrupt} $\Lambda_{\rm s}$CDM model, a one-parameter extension of the standard $\Lambda$CDM model, commonly studied in the literature~\cite{Akarsu:2021fol, Akarsu:2022typ, Akarsu:2023mfb}. This limiting case is expressed as:
\begin{equation}
    \Lambda_{\rm s}(z) \rightarrow \Lambda_{\rm s0}\,{\rm sgn}[z_\dagger - z] \quad \textnormal{for} \quad \eta \rightarrow \infty,
    \label{eq:ssdeff}
\end{equation}
where $\Lambda_{\rm s0} > 0$ represents the present-day value of $\Lambda_{\rm s}(z)$, providing an idealized picture of a rapid AdS-dS transition. While this phenomenological approach within GR has been informative~\cite{Akarsu:2021fol, Akarsu:2022typ, Akarsu:2023mfb}, it lacks a Lagrangian formulation necessary for probing the model's implications on other physical observables, such as solar system constraints and cosmological linear perturbations. To address this limitation, we embed the smooth $\Lambda_{\rm s}$CDM model into $f(T)$ gravity. This approach is advantageous because $f(T)$ gravity, a well-defined extension of the Teleparallel Equivalent of General Relativity (TEGR), allows for a modification of GR through a Lagrangian based on the torsion scalar $T$, rather than the curvature scalar in GR. Embedding the $\Lambda_{\rm s}$CDM framework in $f(T)$ gravity provides a consistent theoretical basis for analyzing the model's impact on both background and perturbative levels. This embedding enables a comprehensive assessment of the model's viability against a wider range of cosmological and astrophysical observations, bridging the gap between theoretical proposals and empirical tests. In this work, we consider a smooth $\Lambda_{\rm s}$CDM model (implied by finite $\eta$) that alters the Hubble parameter $H(z)$ of the $\Lambda$CDM model by replacing the constant $\Lambda$ with the following functional form for $\Lambda_{\rm s}(a)$:
\begin{equation}
    \Lambda_{\rm s}(a) = \Lambda_{\rm dS} \tanh[\zeta(a/a_\dagger - 1)],
    \label{eq:vcdm1}
\end{equation}
where we set $\zeta = 10^{1.5}$ to model a fast transition that closely approximates the background evolution of the abrupt $\Lambda_{\rm s}$CDM model. This approach retains the same number of free parameters as the abrupt $\Lambda_{\rm s}$CDM model, with only one additional parameter, $z_\dagger$, defining the AdS-dS transition redshift, compared to the standard $\Lambda$CDM model\footnote{Larger finite values of $\zeta$ are theoretically possible but would be indistinguishable with current cosmological data. Additionally, for $\zeta = \eta(1+z)$, Eq.~\ref{eq:vcdm1} aligns with $\Lambda_{\rm s}(z) = \Lambda_{\rm dS} \tanh[\eta(z_{\dagger}-z)]$. Since both $\eta$ and $\zeta$ are relevant around $z \sim z_\dagger$ for rapid transitions, this transformation is effectively a scaling, $\zeta \approx \eta(1+z_\dagger)$. Here, we assume $\eta$ is fixed, as in~\cite{Akarsu:2024eoo}.}. By embedding this smooth $\Lambda_{\rm s}$CDM background into $f(T)$ gravity, we provide a model with a Lagrangian formulation that facilitates deeper exploration of its theoretical and observational properties.

The action for generalized teleparallel gravity can be expressed as 
\begin{equation}
    \mathcal{S}=\frac{1}{2\kappa^2}\int d^{4}x\, ||e||f(T)+\mathcal{S}_{M},
    \label{action}
\end{equation}
where $T$ denotes the torsion scalar, $||e||=\det\left({e}_\mu{^a}\right)=\sqrt{-g}$ is the determinant of the tetrad (or vierbein) field, and $\kappa^2\equiv 8\pi G$ with $G$ being the Newton's constant. The term $\mathcal{S}_{M}$ represents the action for matter fields, including baryons, cold dark matter (CDM), photons, and neutrinos. We define the generalized teleparallel function as $f(T)= T + F(T)$, where $F(T)$ encapsulates deviations from the Teleparallel Equivalent of General Relativity (TEGR).

We assume that the background geometry of the universe is described by a spatially flat Friedmann-Lema\^{\i}tre-Robertson-Walker (FLRW) metric. Thus, we consider the Cartesian coordinate system ($t;x,y,z$) and the diagonal vierbein:
\begin{equation}
   {e_{\mu}}^{a}=\textmd{diag}\left[1,a(t),a(t),a(t)\right],
\label{tetrad}
\end{equation}
where $a(t)$ is the scale factor and $t$ is the cosmic time. This vierbein generates the spatially flat FLRW spacetime metric: 
\begin{equation}
   ds^2=dt^{2}-a(t)^{2}\delta_{ij} dx^{i} dx^{j}.
\label{FRW-metric}
\end{equation}
The teleparallel torsion scalar is then defined as:
\begin{equation}
T=-6 H^2, \label{eq:Torsion_sc}
\end{equation}
where $H=\dot{a}/a$ is the Hubble parameter, and the overdot denotes differentiation with respect to $t$.

By varying the action with respect to the vierbein, we derive the field equations:
\begin{eqnarray}
\label{eom}
&&\!\!\!\!\!\!\!\!\!\!\!\!\!\!\!
e^{-1}\partial_{\mu}(ee_A^{\rho}S_{\rho}{}^{\mu\nu})(1+F_{T})
 +
e_A^{\rho}S_{\rho}{}^{\mu\nu}\partial_{\mu}({T})F_{TT}
\nonumber\\
&& -(1+F_{T})e_{A}^{\lambda}T^{\rho}{}_{\mu\lambda}S_{\rho}{}^{\nu\mu}+\frac{1}{4} e_ { A
} ^ {
\nu
}[T+F({T})] \nonumber \\
&&= 4\pi Ge_{A}^{\rho}
\left[{\mathcal{T}^{(m)}}_{\rho}{}^{\nu}+{\mathcal{T}^{(r)}}_{\rho}{}^{\nu}\right],
\end{eqnarray}
where $F_{T}=\partial F/\partial T$, $F_{TT}=\partial^{2} F/\partial T^{2}$,
and, ${\mathcal{T}^{(m)}}_{\rho}{}^{\nu}$ and  ${\mathcal{T}^{(r)}}_{\rho}{}^{\nu}$
are the energy-momentum tensors for matter and radiation, respectively.  

Substituting the vierbein \eqref{tetrad} into the field equations \eqref{eom}, we obtain the modified Friedmann equations:
\begin{eqnarray}\label{background1}
&&H^2= \frac{8\pi G}{3}(\rho_ {\rm m}+\rho_{\rm r})
-\frac{F}{6}+\frac{TF_T}{3},\\\label{background2}
&&\dot{H}=-\frac{4\pi G(\rho_{\rm m}+P_{\rm m}+\rho_{\rm r}+P_{\rm r})}{1+F_{T}+2TF_{TT}},
\end{eqnarray}
where $\rho_{\rm m}$ and $\rho_{\rm r}$ are the energy densities of matter and radiation, respectively, and $P_{\rm m}=0$ and $P_{\rm r}=\rho_{\rm r}/3$ are their corresponding pressures.

From the first Friedmann equation \eqref{background1}, we identify that in $f(T)$ cosmology, the modifications introduce an effective dark energy component of gravitational origin. Specifically, we can express the effective dark energy density as:
\begin{eqnarray} \rho_{\rm DE}\equiv\frac{3}{8 \pi G} \left(-\frac{F}{6}+\frac{T F_T}{3} \right). \end{eqnarray}

To connect the model with observations, we introduce:
\begin{eqnarray}
\label{THdef3}
\frac{H^2(z)}{H^2_{0}}=\frac{T(z)}{T_{0}},
\end{eqnarray}
with $T_0\equiv-6H_{0}^{2}$ as the present-day value of the torsion scalar. Henceforth, a subscript zero on any quantity indicates its value at the present time. Furthermore, using the relations $\rho_{\rm m}=\rho_{\rm m0}(1+z)^{3}$, $\rho_{\rm r}=\rho_{\rm r0}(1+z)^{4}$, we rewrite the first Friedmann equation \eqref{background1} in a more observationally useful form~\cite{Nesseris_2013}:
\begin{eqnarray}
\label{Mod1Ez}
\frac{H^2(z,{\bf r})}{H^2_{0}}=\Omega_{\rm m0}(1+z)^3+\Omega_{\rm r0}(1+z)^4+\Omega_{\rm F0} y(z,{\bf r}),
\end{eqnarray}
where $\Omega_{\rm F0}=1-\Omega_{\rm m0}-\Omega_{\rm r0}$, with $\Omega_{\rm i0}=\frac{8\pi G \rho_{\rm i0}}{3H_0^2}$ as the present-day density parameters. The function  $y(z,{\bf r})$, normalized to unity at the present time, encodes the effects of $f(T)$ modifications and depends on the $f(T)$ functional form parameters $r_1,r_2,...$~\cite{Nesseris_2013}:
\begin{equation}
\label{distortparam}
 y(z,{\bf r})=\frac{1}{T_0\Omega_{\rm F0}}\left(F-2TF_T\right).
\end{equation}
\\

In this work, we introduce a new parametric form for the function $f(T)$, designed to reproduce the background evolution of a smooth $\Lambda_{\rm s}$CDM model while allowing for deviations at the level of linear perturbations.  While $\Lambda_{\rm s}$CDM models typically conform to General Relativity (GR) regarding structure formation, especially in the evolution of the matter density contrast $\delta_{\rm m}$, they do not inherently predict deviations from GR on these scales. In contrast, we propose that such deviations can naturally arise within a modified gravity framework represented by $ f(T) $. To explore this possibility, we consider the following functional form:
\begin{equation}
\label{new_fT}
F(T) = T_0 \Omega_{\rm F0} \tanh \left[ \zeta \left( \frac{a}{a_\dagger} - 1 \right) \right] + \alpha \sqrt{-T}.
\end{equation}

Substituting \eqref{new_fT} into \eqref{distortparam}, we find that the expansion rate of the universe, represented by the function $H(z)$, at the background level, remains approximately consistent with the abrupt $\Lambda_{\rm s}$CDM scenario~\cite{Akarsu:2021fol, Akarsu:2022typ, Akarsu:2023mfb}, as we model the AdS-to-dS transition using $\zeta = 10^{1.5}$ for a rapid transition. At the background level, it also matches exactly with the smooth $\Lambda_{\rm s}$VCDM model~\cite{Akarsu:2024qsi, Akarsu:2024eoo} (a smooth $\Lambda_{\rm s}$CDM framework embedded in type II minimally modified gravity, known as VCDM) that was observationally examined in~\cite{Akarsu:2024eoo}. However, the smooth $f(T)$-$\Lambda_{\rm s}$CDM model considered here deviates from these models at the level of linear perturbations.

In the context of $ f(T) $ gravity, the equation for linear matter perturbations in the subhorizon limit can be expressed as~\cite{Briffa_2024}:
\begin{equation} 
\label{delta_m}
\ddot{\delta}_m + 2H \dot{\delta}_m = 4 \pi G_{\text{eff}} \rho \delta_m, 
\end{equation}
where $G_{\text{eff}}$ denotes the effective Newton's constant, which typically depends on both the redshift $z$ and the cosmic wave vector $k$. For the specific limits and datasets considered in this work, $G_{\text{eff}}$ can be approximated as independent of $k$.  Accordingly, to facilitate analysis, we introduce the linear growth function $D(a)$, defined as:
\begin{equation}
D(a) = \frac{\delta(a)}{\delta(1)},
\end{equation}
$D(a)$ is normalized such that $ D(1) = 1$, representing its present-day value. By employing conventional non-relativistic perturbation theory, we can rewrite \eqref{delta_m} in terms of conformal time ($\eta$) as:
\begin{equation}
D'' + {\cal H} D' - D \left( \frac{3}{2} \frac{G_{\rm eff}}{G} a^2 \rho \right) = 0,
\end{equation}
where a prime denotes a derivative with respect to conformal time, ${\cal H}$ is the Hubble parameter in conformal time units, and $\rho$ represents the total matter density. Using the quasi-static approximation and the modified Poisson equation in the context of $f(T)$ gravity, we have \cite{Golovnev_2018, Nunes_2018}:
\begin{equation} 
\frac{G_{\text{eff}}(z)}{G} = \frac{1}{1 + F_T},
\end{equation}
where $G_{\text{eff}}(z)$ is the effective Newton's constant that governs perturbations, while $G$ is the standard Newtonian constant that governs the background dynamics, such as $H(z)$. 

In our model, characterized by \eqref{new_fT}, this expression becomes:

\begin{equation}
\label{GeffGH}
\frac{G_{\text{eff}}}{G} = \frac{1}{1 + \frac{\alpha}{2 \sqrt{6} H}},
\end{equation}
indicating that the parameter $\alpha$ directly modulates the gravitational strength, thereby altering structure growth predictions compared to the standard $\Lambda$CDM and $\Lambda_{\rm s}$CDM scenarios, based on GR, in a specific manner.
Substituting \eqref{GeffGH} into the modified perturbation equation and using $\mathcal{H}=aH$, we obtain:
Substituting \eqref{GeffGH} into the modified perturbation equation, in conformal time units, we obtain:
\begin{equation}
\label{perteff}
D'' + \mathcal{H} D' - \frac{3}{2} a^2 \rho_{\rm m} D \left(1 - \frac{\alpha a}{2 \sqrt{6} \mathcal{H}}\right) = 0,
\end{equation} 
for small values of $\alpha$ (viz., $\alpha \ll2\sqrt{6}\mathcal{H}\,a^{-1}$). From this equation, it is clear that if $\alpha > 0$, the effective gravitational strength is reduced, resulting in a suppression of the growth of perturbations. Conversely, if $\alpha < 0$, the effective gravitational strength is enhanced, leading to an increase in the growth rate of perturbations.

Thus, the $f(T)$-$\Lambda_{\rm s}$CDM model under consideration, as defined in \eqref{new_fT}, can be regarded as effectively equivalent to the widely studied phenomenological abrupt $\Lambda_{\rm s}$CDM model, which assumes GR, at the background level when a fast AdS-to-dS transition epoch is assumed. This makes our $f(T)$-$\Lambda_{\rm s}$CDM model nearly indistinguishable from the abrupt $\Lambda_{\rm s}$CDM model based on current background-level observations. Additionally, similar to the abrupt $\Lambda_{\rm s}$CDM model, our $f(T)$-$\Lambda_{\rm s}$CDM model matches the standard $\Lambda$CDM model's expansion rate, $H(z)$, after the transition at $z < z_\dagger$ ($a > a_\dagger$). However, as shown by \eqref{perteff}, differences arise at the linear perturbation level, governed by the parameter $\alpha$, which modifies the effective gravitational strength. When $\alpha = 0$, the $f(T)$-$\Lambda_{\rm s}$CDM and abrupt $\Lambda_{\rm s}$CDM models become equivalent at both the background and perturbative levels. Since these deviations manifest solely at the perturbative level, $\alpha$ can only be constrained through CMB data and structure formation observations. In this work, we specifically focus on using RSD (redshift-space distortions) data to probe these perturbative differences and the influence of $\alpha$ on the growth of cosmic structures.

With the main equations outlined, we now turn our attention to how structure formation is affected, focusing particularly on linear scales. To efficiently assess the impact on the evolution of matter perturbations, it is essential to compare theoretical predictions with cosmological observables, such as redshift-space distortions (RSD). These distortions result from velocity-induced effects that arise when mapping from real space to redshift space, due to the peculiar motions of objects along the line of sight. Such distortions introduce anisotropies in the observed clustering patterns, which are directly influenced by the growth of cosmic structures. RSD measurements are sensitive to the combination$f\sigma_8(z)$, or equivalently$f(a)\sigma_8(a)$, where$\sigma_8(a)$represents the variance of the mass distribution smoothed on a sphere of radius$R = 8h^{-1} \text{Mpc}$, and$f(a)$is the logarithmic derivative of the linear growth function$D(a) = \delta_m(a)/\delta_m(1)$with respect to the scale factor:
\begin{equation}
f(a) \equiv \frac{d \ln D(a)}{d \ln a},
\end{equation}
where the matter density perturbation$\delta_{\rm m}$is given by$\rho_{\rm m} \delta_{\rm m} = \rho_{\rm b} \delta_{\rm b} + \rho_{\rm c} \delta_{\rm c}$, representing the combined contributions from baryonic and cold dark matter.

Given these new properties, we refer to this class of models as $f(T)-\Lambda_{\rm s}$CDM gravity models. In the following sections, we will explore the new observational constraints for this scenario.

\section{Data and Methodology}
\label{data}

To derive constraints on the model baseline, we utilize the following datasets:
\begin{itemize}
\item \textit{Redshift Space Distortions} (\textbf{RSD}): Numerous measurements of $f\sigma_8(z)$ from various surveys are documented in the literature, each involving different assumptions and subject to distinct uncertainties. Before incorporating any of these measurements, it is essential to assess their internal consistency. Such an evaluation is undertaken using a Bayesian model comparison framework, as detailed in Ref.~\cite{Sagredo_2018}. This framework includes a comprehensive analysis of the $f(z)\sigma_8(z)$ measurements listed in Table I of~\cite{Sagredo_2018}, encompassing 22 data points spanning the redshift range $0.02 < z < 1.944$. We refer to this dataset as \texttt{RSD}.\\

\item \textit{Cosmic Chronometers} (\textbf{CC}):
Measurements of the expansion rate $H(z)$ derived from the relative ages of massive, early-time, passively-evolving galaxies, known as Cosmic Chronometers~\cite{Jimenez:2001gg}. In our analyses, we conservatively use only a compilation of 15 CC measurements in the redshift range $0.179\lesssim z\lesssim 1.965$~\cite{Moresco:2012by,Moresco:2015cya,Moresco:2016mzx}, accounting for all non-diagonal terms in the covariance matrix and systematic contributions. We refer to this dataset as \texttt{CC}.\\

\item \textit{Baryon Acoustic Oscillations} (\textbf{DESI-BAO}): 
Baryon Acoustic Oscillations (BAO) measurements provided by Dark Energy Spectroscopic Instrument (DESI) collaboration from observations of galaxies and quasars~\cite{desicollaboration2024desi2024iiibaryon}, and Lyman-$\alpha$ tracers~\cite{desicollaboration2024desi2024ivbaryon}, as summarized in Table~I of Ref.~\cite{desicollaboration2024desi2024vicosmological}. These measurements consist of both isotropic and anisotropic BAO data in the redshift range $0.1 < z < 4.2$ and are divided into seven redshift bins. The isotropic BAO measurements are represented as $D_{\mathrm{V}}(z)/r_{\mathrm{d}}$, where $D_{\mathrm{V}}$ denotes the angle-averaged distance, normalized to the (comoving) sound horizon at the drag epoch. The anisotropic BAO measurements include $D_{\mathrm{M}}(z)/r_{\mathrm{d}}$ and $D_{\mathrm{H}}(z)/r_{\mathrm{d}}$, where $D_{\mathrm{M}}$ is the comoving angular diameter distance and $D_{\mathrm{H}}$ is the Hubble horizon. Additionally, the correlation between the measurements of $D_{\mathrm{M}}/r_{\mathrm{d}}$ and $D_{\mathrm{V}}/r_{\mathrm{d}}$ is also taken into account. We refer to this dataset as \texttt{DESI}. \\

\item \textit{Type Ia Supernovae} (\textbf{SN Ia}): Type Ia supernovae act as standardizable candles, providing a crucial method for measuring the universe's expansion history and supporting $\Lambda$-dominated models. In this work, we use the following recent samples:
\begin{enumerate}

    \item [(i)] \textbf{PantheonPlus}: We incorporated SN Ia distance modulus measurements from the PantheonPlus sample~\cite{Brout:2022vxf}, which consists of 1550 supernovae spanning a redshift range from 0.01 to 2.26. We refer to this dataset as \texttt{PP}.

    \item [(ii)] \textbf{Union 3.0}: The Union 3.0 compilation, consisting of 2087 SN Ia, was presented in~\cite{Brout:2022vxf}. Notably, 1363 of these SN Ia are common with the PantheonPlus sample. This dataset features a distinct treatment of systematic errors and uncertainties, employing Bayesian Hierarchical Modeling. We refer to this dataset as \texttt{Union3}.

    \item [(iii)] \textbf{DESY5}: As part of their Year 5 data release, the Dark Energy Survey (DES) recently published results from a new, homogeneously selected sample of 1635 photometrically classified SN Ia with redshifts spanning $0.1 < z < 1.3$~\cite{descollaboration2024darkenergysurveycosmology}. This sample is complemented by 194 low-redshift SN Ia (shared with the PantheonPlus sample) in the range $0.025 < z < 0.1$. We refer to this dataset as \texttt{DESY5}.\\
\end{enumerate}

\end{itemize}

In all analyses presented in this work, we incorporate state-of-the-art Big Bang Nucleosynthesis  (BBN) data, comprising of measurements of the primordial abundances of helium $Y_P$ \cite{Aver:2015iza} and the deuterium measurement $y_{DP} = 10^5 n_D/n_H$ \cite{Cooke:2017cwo}. It is known that the BBN likelihood is sensitive to constraints on the physical baryon density $\omega_b \equiv \Omega_bh^2$ and the effective number of neutrino species $N_{\rm eff}$. We fix $N_{\rm eff}= 3.046$ in the present work. For theoretical predictions, we use the baseline likelihood provided by the \texttt{PArthENoPE 2.0} code~\cite{Consiglio:2017pot}.

All cosmological observables are computed with \texttt{CLASS}~\cite{Diego_Blas_2011} and \texttt{MontePython}~\cite{brinckmann2018montepython3boostedmcmc, Benjamin_Audren_2013}. We assess the convergence of the Markov Chain Monte Carlo (MCMC) chains using the Gelman-Rubin parameter $R-1$, requiring $ R-1 < 0.01 $ for convergence. In our analyses, we assume flat priors for all baseline parameters with wide ranges: $\omega_b \in [0.0, 1.0]$, $\omega_{\text{cdm}} \in [0.0, 1.0]$, $\sigma_8  \in [0.2, 2.0]$, $z_{\dagger} \in [1.0, 5.0]$ and $\alpha \in [0, 1]$. Values of $\alpha$ are assumed to be positive to ensure the stability of model variations, maintaining a positive effective Newton's constant throughout the evolution of the matter density contrast. In the following sections, we present and discuss our main results.

\section{Main results and discussions}
\label{results}

Table~\ref{tab_R1} summarizes our statistical results for the $f(T)$-$\Lambda_{\rm s}$CDM model, considering only geometric measurements. In other words, it does not include the effects of the evolution of matter density contrasts, i.e., RSD measurements. Essentially, these results represent a revision of the discussions recently presented in~\cite{Akarsu:2024eoo}. As a novel aspect, we present updated results incorporating the latest SNe Ia data from the DESY5 and Union 3.0 compilations. It is important to emphasize that none of the results discussed here incorporate CMB data. The inclusion of BAO+BBN in our analyses provides constraining power similar to that of CMB data, as these scenarios predict only theoretical changes in the Hubble parameter $H(z)$. However, similar constraint strength does not imply identical correlations within the parameter space of the baseline model. Therefore, the potential benefits of incorporating CMB data will be addressed in a future study, as the perturbative developments in the context of $f(T)$ modified gravity are still underway.
\begin{table*}[htpb!]
    \centering\caption{Marginalized constraints and mean values with 68\% CL on the free and some derived parameters of the $\Lambda_{\rm s}$CDM model from the combinations of the DESI, PP, Union3, DES5Y, and CC datasets.}
    \renewcommand{\arraystretch}{1.3}
    \begin{tabular}{l c c c c}
        \hline
        Dataset  &  PP+CC+DESI   &  Union3+CC+DESI  &  DES5Y+CC+DESI \\
        \hline
        $\Omega_{\rm m}$ & $0.309^{+0.012}_{-0.011}$  & $0.307^{+0.013}_{-0.012}$ & $0.3232^{+0.0094}_{-0.0071}$ \\
        $H_0$ & $69.60\pm0.75$  & $69.58\pm 0.75$ & $69.62\pm0.68$ \\
        $z_{\dagger}$ & $2.96^{+0.46}_{-0.64}$ & $2.97^{+0.45}_{-0.64}$ & $2.77^{+0.74}_{-0.55}$ \\
        \hline
        \hline
    \end{tabular}
    \label{tab_R1}
\end{table*}
\begin{table*}[htpb!]
    \centering\caption{Marginalized constraints and mean values with 68\% confidence levels (CL) on the free and some derived parameters of the $f(T)$-$\Lambda_{\rm s}$CDM framework, assuming a fixed $\alpha = 0$, from the combinations of the RSD, DESI,  PP, Union3, DES5Y, and CC datasets.}
    \renewcommand{\arraystretch}{1.3}
    \begin{tabular}{l c c c c}
        \hline
Dataset & PP+CC+DESI+RSD & Union3+CC+DESI+RSD & DESY5+CC+DESI+RSD \\
\hline
$\Omega_{\mathrm{m}}$   &  $0.299\pm 0.011$   &  $0.306\pm 0.012$              &  $0.322^{+0.010}_{-0.0072}$      \\

$H_0$   &   $67.93\pm 0.16$     &  $69.58\pm 0.74$              &  $69.61\pm 0.69$        \\
$\sigma_8$      &    $0.759\pm 0.028$      &  $0.753\pm 0.027$              &  $0.740\pm 0.028$      \\
$S_8$   &    $0.758\pm 0.027$      &   $0.761\pm 0.027$              &   $0.767\pm 0.028$      \\
$z_\dagger$   &   $3.09^{+0.32}_{-0.65}$     &  $2.99^{+0.43}_{-0.65}$        &  $2.73^{+0.84}_{-0.51}$         \\
        \hline
        \hline
    \end{tabular}
\label{tab_R2}
\end{table*}

\begin{table*}[htpb!]
    \centering\caption{Marginalized constraints and mean values with 68\% CL on the free and some derived parameters of the $f(T)$-$\Lambda_{\rm s}$CDM framework, assuming $\alpha$ as a free parameter, from the combinations of the RSD, DESI, PP, Union3, DES5Y, and CC datasets.}
    \renewcommand{\arraystretch}{1.3}
    \begin{tabular}{lcccc}
        \hline
        Dataset  &    PP+CC+DESI+RSD  &  Union3+CC+DESI+RSD  &  DES5Y+CC+DESI+RSD \\
        \hline
        $\Omega_{\rm m}$ &  $0.311^{+0.012}_{-0.010}$ & $0.310^{+0.012}_{-0.011}$ & $0.320\pm 0.010$ \\
        
        $H_0$ &  $69.56\pm 0.74$ & $69.80\pm 0.70$ & $68.56^{+0.49}_{-0.32}$ \\
        
        $\sigma_8$ &  $0.884\pm 0.034$ & $0.876^{+0.027}_{-0.022}$ & $0.839\pm 0.017$ \\

        $S_8$ &  $0.900^{+0.050}_{-0.045}$ &  $0.890^{+0.043}_{-0.036}$ &  $0.867\pm 0.031$ \\
        
        $\alpha$ &  $0.00073^{+0.00027}_{-0.00033}$ & $0.00068^{+0.00024}_{-0.00029}$ & $0.00052^{+0.00019}_{-0.00024}$ \\
        
        $z_{\dagger}$ &  $2.538^{+0.099}_{-0.25}$ & $2.65^{+0.25}_{-0.29}$ & $2.94^{+0.36}_{-0.60}$ \\
        \hline
        \hline
    \end{tabular}
    
    \label{tab_R3}
\end{table*}

\begin{figure*}[htpb!]
    \centering
    \includegraphics[width=0.40\linewidth]{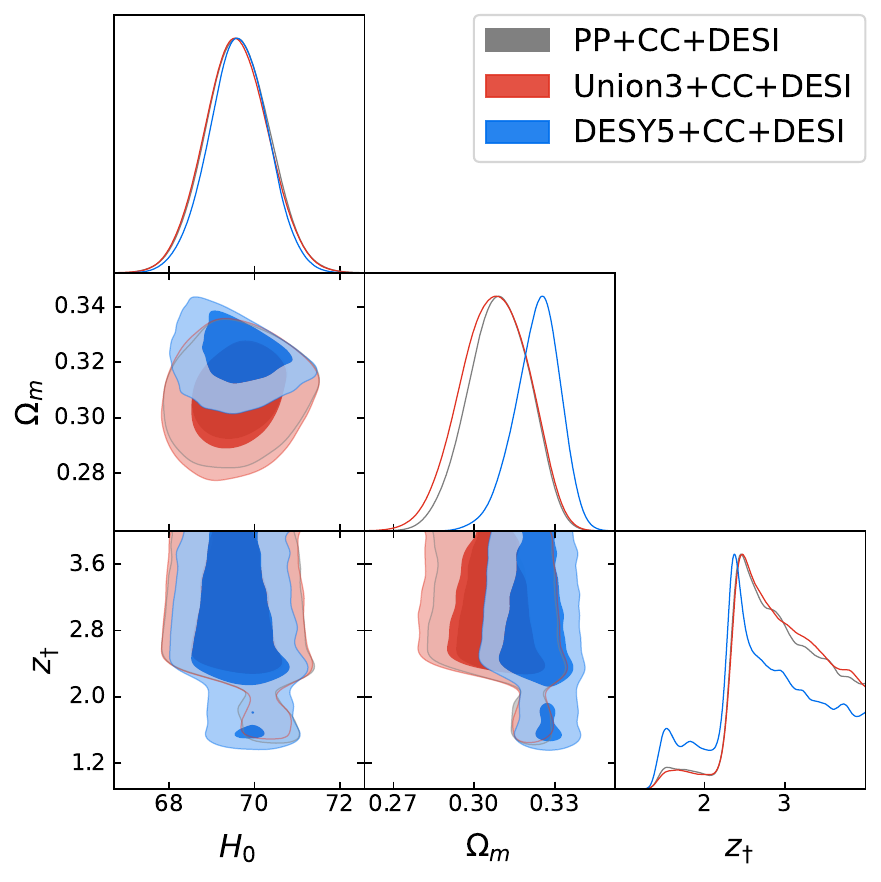} \,\,\,
        \includegraphics[width=0.57\linewidth]{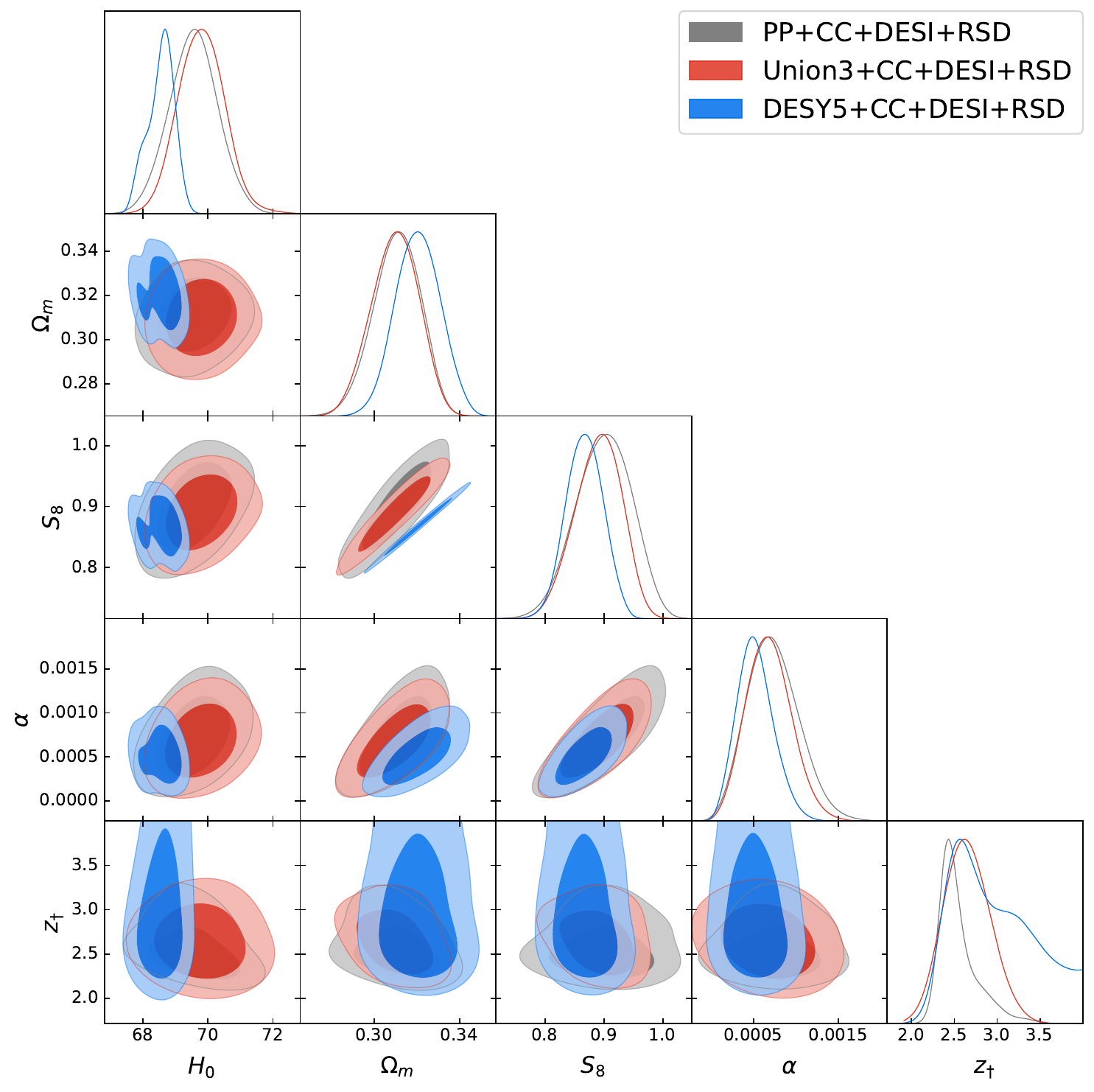}
    \caption{Marginalized posterior distributions and 68\% and 95\% CL contours for some selected parameters of $\Lambda_{\rm s}$CDM model (left panel) and $f(T)$-$\Lambda_{\rm s}$CDM model (right panel) from different combinations of datasets, as indicated in the legends.}
    \label{fig:PS_1}
\end{figure*}

We begin by interpreting the redshift of transition from a combined analysis of PP, CC, and DESI data. In this case, we find $z_{\dagger} = 2.96^{+0.46}_{-0.64}$ at 68\% CL, which is highly consistent with previous estimates~\cite{Akarsu:2021fol,Akarsu:2022typ,Akarsu:2023mfb,Akarsu:2024eoo}. It is also important to highlight that our analysis includes DESI data, which were not present in earlier studies. Following this, we consider joint analyses of PP+CC+DESI, Union3+CC+DESI, and DESY5+CC+DESI, and we observe similarly robust observational constraints on the parameter $z_{\dagger}$. As previously discussed in~\cite{Akarsu:2024eoo}, the inclusion of BAO data tends to keep the values of $H_0$ lower compared to local measurements inferred by the SH0ES team~\cite{Murakami:2023xuy, Riess:2021jrx}. Figure~\ref{fig:PS_1} (left panel) presents the marginalized one- and two-dimensional distributions (68\% and 95\% CL) for the $\Lambda_{\rm s}$CDM model, derived from geometric data analyses. This reanalysis incorporates updated DESI data and the latest SNe Ia samples from the DESY5 and Union 3.0 compilations, offering refined insights into the model. The estimates for $z_{\dagger}$ remain consistent with previous results~\cite{Akarsu:2021fol,Akarsu:2022typ,Akarsu:2023mfb,Akarsu:2024eoo}.

Table~\ref{tab_R2} includes RSD data, but only considering effects due to changes in the Hubble parameter, i.e., via a modified $H(z)$ function. In other words, we assume $\alpha = 0$ in all analyses quantified in this table. From the perspective of the joint analysis PP+DESI+CC+RSD, we find $z_{\dagger} = 2.87^{+0.72}_{-0.53}$, which can be compared to $z_{\dagger} = 2.96^{+0.46}_{-0.64}$ without RSD data. That is, we observe a small gain in precision in the constraints, but the results remain consistent with each other. We interpret a similar trend for the other analyses. On the other hand, the inclusion of RSD now allows us to constrain the $S_8$ parameter within this new scenario. In general, we observe that $S_8$ tends to remain at lower values across all analyses (see Table~\ref{tab_R2}). This trend is expected, as any effects on the growth function are not being accounted for; instead, we are only considering modifications in the $H(z)$ function.

In Table~\ref{tab_R3}, we present the results of our constraints, considering all theoretical corrections predicted by the $f(T)$-$\Lambda_{\rm s}$CDM scenario. Specifically, we include the presence of a transition, $z_{\dagger}$, while also treating $\alpha$ as a free parameter. The transition $z_{\dagger}$ remains consistent with all previously discussed cases, but significant impacts are now observed in this scenario. The potential change in the growth function induced by $f(T)$ gravity through the parameter $\alpha$ introduces new correlations between the parameters $H_0$ and $S_8$ (see Figure~\ref{fig:PS_1}). More specifically, $\alpha$ exhibits a positive correlation with both $S_8$ and $H_0$, naturally leading to higher values for both parameters.

\begin{figure}[htpb!]
    \centering
    \includegraphics[width=1.0\linewidth]{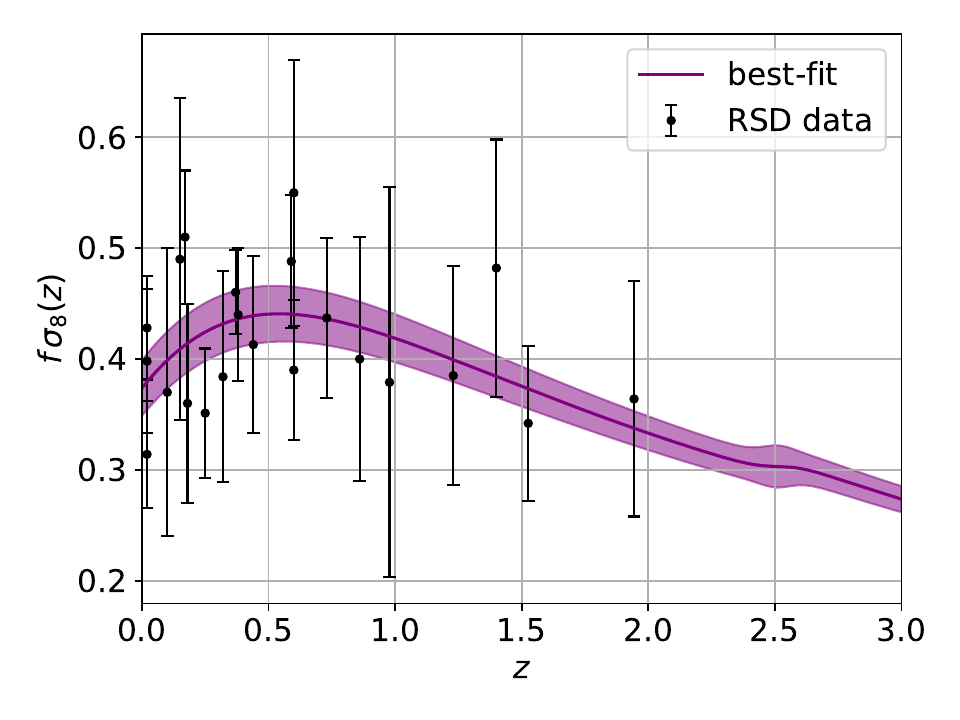}
    \caption{Statistical reconstruction of the theoretical prediction $f\sigma_8(z)$ at 2$\sigma$ confidence levels for the $f(T)$-$\Lambda_{\rm s}$CDM model through the joint analysis of PP+CC+DESI+RSD, compared to RSD measurements.}
    \label{fs8_rec}
\end{figure}

As previously discussed, due to the presence of BAO data, even with the introduction of a new positive correlation between $\alpha$ and $H_0$, the values of $H_0$ remain insufficient to resolve the $H_0$ tension. On the other hand, the new correlation in the $\alpha$-$S_8$ plane significantly strengthens the constraints on the $S_8$ parameter. It is important to note that, in the absence of $\alpha$, $S_8$ exhibits lower values, as expected (see Table~\ref{tab_R2}). From the perspective of interpreting a potential tension in $S_8$, this suggests that this new class of $f(T)$-$\Lambda_{\rm s}$CDM models has the potential to resolve the $S_8$ tension by increasing its value, thus making it compatible with CMB measurements. Typically, this problem is approached in the opposite direction in the literature, with models proposed to lower the expected values from CMB to match the lower $S_8$ measurements from Large Scale Structure observations.

Another interesting point is that the data show a significant preference for $\alpha \neq 0$ in all analyses conducted. For the combined analysis of PP+CC+DESI+RSD, we find $\alpha = 0.00073^{+0.00027}_{-0.00033}$ at 68\% CL. This joint analysis provides evidence for $\alpha>0$ at more than 2$\sigma$ CL. We observe a similar trend in the Union3+CC+DESI+RSD and DES5Y+CC+DESI+RSD analyses. 

Thus, by considering linear perturbative effects not predicted in the standard $\Lambda$CDM and $\Lambda_{\rm s}$CDM models based on GR, we identify a significant preference for modifications in the growth function of structures. Figure~\ref{fs8_rec} presents a statistical reconstruction of the observable $f\sigma_8(z)$ at 2$\sigma$ CL, along with the best-fit prediction from the combined PP+CC+DESI+RSD analysis.

\begin{figure}[htpb!]
    \centering
    \includegraphics[width=1.0\linewidth]{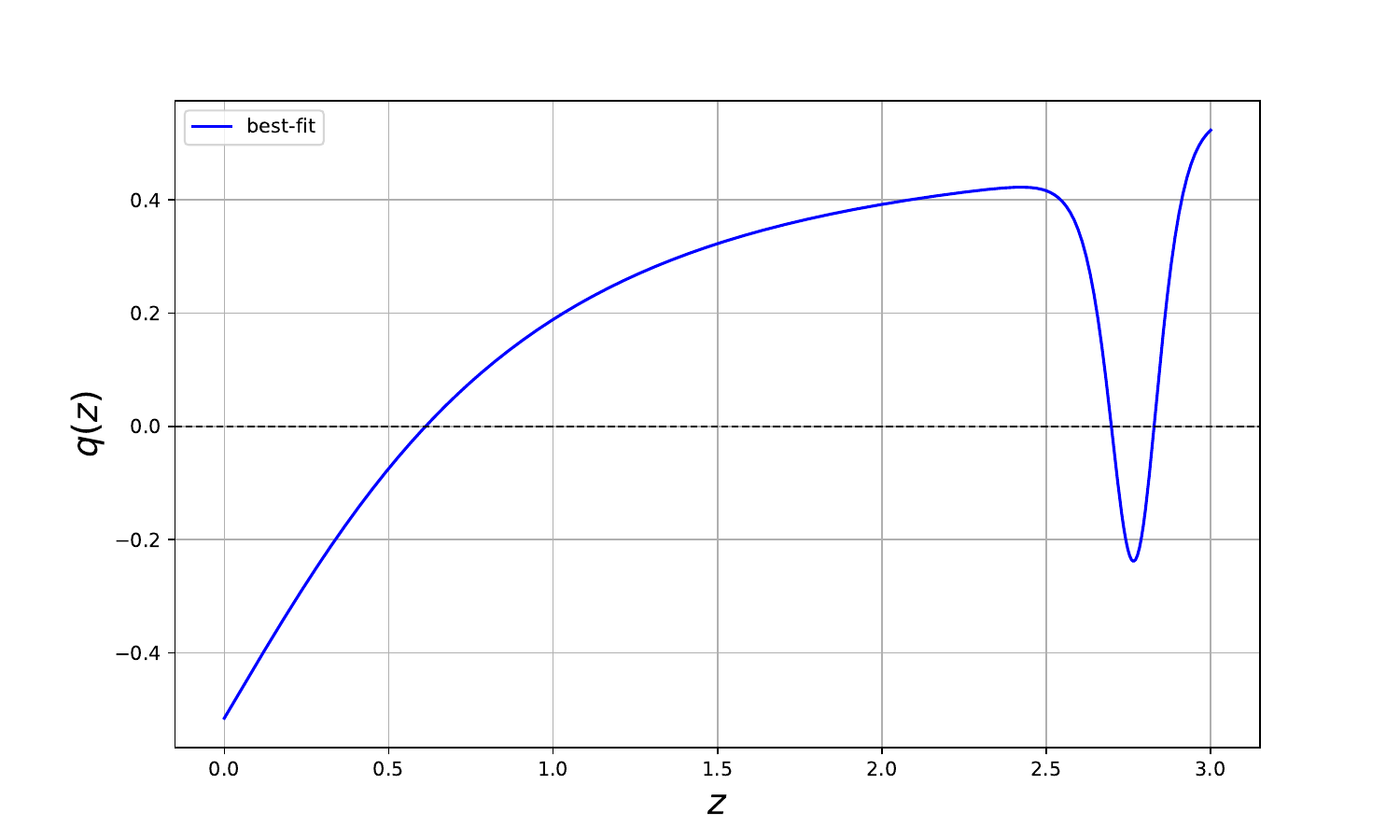}
    \caption{Statistical reconstruction of the theoretical prediction $q(z)$ for the $f(T)$-$\Lambda_{\rm s}$CDM model using best-fit values obtained from the joint analysis of DES5Y, CC, and DESI data.}
    \label{q_rec}
\end{figure}
Figure \ref{q_rec} presents a theoretical reconstruction of the deceleration parameter $q(z)$ using the joint analysis of DES5Y+CC+DESI data. This combination provides comparable constraining power to any other dataset considered in this work for background-level parameter inference, effectively constraining the parameters necessary to predict $q(z)$. From $z=0$ to $z=2$, the behavior of $q(z)$ follows the expected trend, aligning well with the predictions of the standard $\Lambda$CDM model, including the transition from decelerated to accelerated expansion at $z \sim 0.55$. Beyond $z=3$, $q$ approaches 0.5, consistent with the standard cosmological model during a matter-dominated universe. As first suggested in Ref.~\cite{Akarsu:2024eoo}, our model predicts an additional, temporary phase of accelerated expansion occurring shortly after the AdS-dS transition begins. This phase emerges when the effective dark energy density becomes positive with an equation of state (EoS) less than $-1$, triggering a brief period of accelerated cosmic expansion lasting for $\Delta z \sim 0.15$ around $z \sim 2.7$. Subsequently, the universe exits this temporary accelerated expansion phase and gradually approaches the behavior predicted by the $\Lambda$CDM model near $z \sim 2.5$, corresponding to a stage before the present-day accelerated expansion begins at $z \sim 0.55$. This distinct feature in the behavior of $q(z)$, induced by the rapid dynamics associated with a mirror AdS-dS transition, represents a novel prediction of the smooth $\Lambda_{\text{s}}$CDM model. Its potential to serve as smoking-gun evidence or to falsify the model highlights the theoretical richness of this cosmological framework. Since no direct observational data currently exist at $z \sim 2.7$, this prediction strongly motivates future exploration of this redshift range in cosmological studies.


\section{Final Remarks}
\label{final}
The concept of a rapidly sign-switching cosmological constant, interpreted as a mirror AdS-dS transition in the late universe at $z\sim2$ and known as the $\Lambda_{\rm s}$CDM, has significantly improved the fit to observational data, offering a promising framework for alleviating major cosmological tensions such as the $H_0$ and $S_8$ tensions~\cite{Akarsu:2021fol,Akarsu:2022typ,Akarsu:2023mfb,Akarsu:2024qsi,Akarsu:2024eoo,Yadav:2024duq}. Within the standard framework of general relativity (GR), these models predict alterations in the universe's expansion rate exclusively through modifications to the Hubble parameter $H(z)$, without influencing the rate of structure formation beyond what is expected from GR.

Conversely, the processes of structure formation and evolution provide crucial astrophysical and cosmological insights into the dark sector of the universe and may even hint at modifications to General Relativity (GR). In this work, we propose a new cosmological model that generalizes the frequently studied $ \Lambda_{\rm s} $CDM model within the GR framework by introducing a novel phenomenological parametrization within the $ f(T) $ gravity framework. Dubbed the $ f(T) $-$ \Lambda_{\rm s} $CDM model, this framework remains indistinguishable from the standard $ \Lambda_{\rm s} $CDM model based on GR at the background level but exhibits different behavior at the level of linear perturbations, which has significant implications for structure formation and cosmological observations.

The key results and contributions presented in this work can be summarized as follows:

\begin{itemize}
    \item We update the observational constraints within the context of the $\Lambda_{\rm s}$CDM framework using the latest BAO-DESI and SNe Ia measurements, incorporating the recent DESY5 and Union3 compilations. The AdS-to-dS transition redshift $z_{\dagger}$ is found to be compatible with previous results reported in the literature.\\

    \item We introduce a novel gravitational model within the framework of $ f(T) $ gravity that remains indistinguishable from the standard GR-based $ \Lambda_{\rm s} $CDM model at the background cosmological level but predicts differences in the growth rate of structures. A new degree of freedom, $ \alpha $, is introduced to quantify these perturbative effects.\\

    \item We apply RSD data for the first time in both the context of the $ \Lambda_{\rm s} $CDM model and the newly proposed $ f(T) $-$ \Lambda_{\rm s} $CDM model in this work. With the inclusion of RSD data, we find that $ \alpha > 0 $ at more than $ 2\sigma $ confidence level (CL), suggesting that this model fits the data better than the standard $ \Lambda $CDM model.\\

    \item Due to a new positive correlation in the $ \alpha $-$ S_8 $ plane, this scenario has the potential to resolve the current observational $ S_8 $ tension identified in Large Scale Structure observations.
\\
\end{itemize}

In conclusion, the $f(T) $-$\Lambda_{\rm s} $CDM model proposed in this work successfully implements the $\Lambda_{\rm s} $CDM scenario within teleparallel $f(T) $ modified gravity by introducing a new degree of freedom through the parameter $\alpha $. This parameter alters the growth rate of cosmic structures without affecting the background cosmological evolution. Our results, particularly the finding that $\alpha > 0 $ at more than $2\sigma $ CL using RSD data, indicate that this new model provides a better fit to current observational datasets, including BAO and SNe Ia. Furthermore, the positive correlation between $\alpha $ and $S_8 $ suggests that this model has the potential to resolve the so-called $S_8 $ tension identified in LSS observations. Specifically, this model predicts higher values for $S_8 $, making them more compatible with CMB data.

Future work will focus on extending our analysis by incorporating CMB data, which is not included in the present study. The perturbative effects of $f(T)$ gravity on CMB anisotropies, particularly concerning linear growth and structure formation, remain an open question. Including CMB data will provide a more comprehensive test of the $f(T)$-$\Lambda_{\rm s}$CDM model and help clarify its potential for resolving the $H_0$ and $S_8$ tensions in a consistent manner. As the framework of $f(T)$ gravity continues to develop, further investigations into non-linear effects and their implications for the late-time universe will also be crucial. These efforts will pave the way for a more robust understanding of modified gravity theories and their role in the evolution and dynamics of the cosmos.

\begin{acknowledgments}
\noindent  M.S. and A.B. received support from the CNPq and CAPES scholarship, respectively. M.S. and A.B. thank Miguel Sabogal for his valuable discussions and assistance in developing the codes and likelihood used in this work. R.C.N. thanks the CNPq under the project No. 304306/2022-3, and FAPERGS for partial financial support under the project No. 23/2551-0000848-3. \"{O}.A. acknowledges the support by the Turkish Academy of Sciences in the scheme of the Outstanding Young Scientist Award  (T\"{U}BA-GEB\.{I}P). This study was supported by Scientific and Technological Research Council of Turkey (TUBITAK) under the Grant Number~122F124. The authors thank TUBITAK for their support. S.K. gratefully acknowledges the support of Startup Research Grant from Plaksha University  (File No. OOR/PU-SRG/2023-24/08). This article/publication is based upon work from COST Action CA21136 Addressing observational tensions in cosmology with systematics and fundamental physics (CosmoVerse) supported by COST (European Cooperation in Science and Technology).
\end{acknowledgments}

\bibliography{main}

\begin{thebibliography}{88}%
\makeatletter
\providecommand \@ifxundefined [1]{%
 \@ifx{#1\undefined}
}%
\providecommand \@ifnum [1]{%
 \ifnum #1\expandafter \@firstoftwo
 \else \expandafter \@secondoftwo
 \fi
}%
\providecommand \@ifx [1]{%
 \ifx #1\expandafter \@firstoftwo
 \else \expandafter \@secondoftwo
 \fi
}%
\providecommand \natexlab [1]{#1}%
\providecommand \enquote  [1]{``#1''}%
\providecommand \bibnamefont  [1]{#1}%
\providecommand \bibfnamefont [1]{#1}%
\providecommand \citenamefont [1]{#1}%
\providecommand \href@noop [0]{\@secondoftwo}%
\providecommand \href [0]{\begingroup \@sanitize@url \@href}%
\providecommand \@href[1]{\@@startlink{#1}\@@href}%
\providecommand \@@href[1]{\endgroup#1\@@endlink}%
\providecommand \@sanitize@url [0]{\catcode `\\12\catcode `\$12\catcode `\&12\catcode `\#12\catcode `\^12\catcode `\_12\catcode `\%12\relax}%
\providecommand \@@startlink[1]{}%
\providecommand \@@endlink[0]{}%
\providecommand \url  [0]{\begingroup\@sanitize@url \@url }%
\providecommand \@url [1]{\endgroup\@href {#1}{\urlprefix }}%
\providecommand \urlprefix  [0]{URL }%
\providecommand \Eprint [0]{\href }%
\providecommand \doibase [0]{http://dx.doi.org/}%
\providecommand \selectlanguage [0]{\@gobble}%
\providecommand \bibinfo  [0]{\@secondoftwo}%
\providecommand \bibfield  [0]{\@secondoftwo}%
\providecommand \translation [1]{[#1]}%
\providecommand \BibitemOpen [0]{}%
\providecommand \bibitemStop [0]{}%
\providecommand \bibitemNoStop [0]{.\EOS\space}%
\providecommand \EOS [0]{\spacefactor3000\relax}%
\providecommand \BibitemShut  [1]{\csname bibitem#1\endcsname}%
\let\auto@bib@innerbib\@empty
\bibitem [{\citenamefont {Clifton}\ \emph {et~al.}(2012)\citenamefont {Clifton}, \citenamefont {Ferreira}, \citenamefont {Padilla},\ and\ \citenamefont {Skordis}}]{Clifton_2012}%
  \BibitemOpen
  \bibfield  {author} {\bibinfo {author} {\bibfnamefont {T.}~\bibnamefont {Clifton}}, \bibinfo {author} {\bibfnamefont {P.~G.}\ \bibnamefont {Ferreira}}, \bibinfo {author} {\bibfnamefont {A.}~\bibnamefont {Padilla}}, \ and\ \bibinfo {author} {\bibfnamefont {C.}~\bibnamefont {Skordis}},\ }\href {\doibase 10.1016/j.physrep.2012.01.001} {\bibfield  {journal} {\bibinfo  {journal} {Physics Reports}\ }\textbf {\bibinfo {volume} {513}},\ \bibinfo {pages} {1–189} (\bibinfo {year} {2012})}\BibitemShut {NoStop}%
\bibitem [{\citenamefont {Ishak}(2018)}]{Ishak_2018}%
  \BibitemOpen
  \bibfield  {author} {\bibinfo {author} {\bibfnamefont {M.}~\bibnamefont {Ishak}},\ }\href {\doibase 10.1007/s41114-018-0017-4} {\bibfield  {journal} {\bibinfo  {journal} {Living Reviews in Relativity}\ }\textbf {\bibinfo {volume} {22}} (\bibinfo {year} {2018}),\ 10.1007/s41114-018-0017-4}\BibitemShut {NoStop}%
\bibitem [{\citenamefont {Nojiri}\ \emph {et~al.}(2017)\citenamefont {Nojiri}, \citenamefont {Odintsov},\ and\ \citenamefont {Oikonomou}}]{Nojiri_2017}%
  \BibitemOpen
  \bibfield  {author} {\bibinfo {author} {\bibfnamefont {S.}~\bibnamefont {Nojiri}}, \bibinfo {author} {\bibfnamefont {S.}~\bibnamefont {Odintsov}}, \ and\ \bibinfo {author} {\bibfnamefont {V.}~\bibnamefont {Oikonomou}},\ }\href {\doibase 10.1016/j.physrep.2017.06.001} {\bibfield  {journal} {\bibinfo  {journal} {Physics Reports}\ }\textbf {\bibinfo {volume} {692}},\ \bibinfo {pages} {1–104} (\bibinfo {year} {2017})}\BibitemShut {NoStop}%
\bibitem [{\citenamefont {Saridakis}\ \emph {et~al.}(2023)\citenamefont {Saridakis}, \citenamefont {Lazkoz} \emph {et~al.}}]{saridakis2023modifiedgravitycosmologyupdate}%
  \BibitemOpen
  \bibfield  {author} {\bibinfo {author} {\bibfnamefont {E.~N.}\ \bibnamefont {Saridakis}}, \bibinfo {author} {\bibfnamefont {R.}~\bibnamefont {Lazkoz}},  \emph {et~al.},\ }\href {https://arxiv.org/abs/2105.12582} {\enquote {\bibinfo {title} {Modified gravity and cosmology: An update by the cantata network},}\ } (\bibinfo {year} {2023}),\ \Eprint {http://arxiv.org/abs/2105.12582} {arXiv:2105.12582 [gr-qc]} \BibitemShut {NoStop}%
\bibitem [{\citenamefont {Frusciante}\ and\ \citenamefont {Perenon}(2020)}]{Frusciante_2020}%
  \BibitemOpen
  \bibfield  {author} {\bibinfo {author} {\bibfnamefont {N.}~\bibnamefont {Frusciante}}\ and\ \bibinfo {author} {\bibfnamefont {L.}~\bibnamefont {Perenon}},\ }\href {\doibase 10.1016/j.physrep.2020.02.004} {\bibfield  {journal} {\bibinfo  {journal} {Physics Reports}\ }\textbf {\bibinfo {volume} {857}},\ \bibinfo {pages} {1–63} (\bibinfo {year} {2020})}\BibitemShut {NoStop}%
\bibitem [{\citenamefont {Maluf}(2013)}]{Maluf_2013}%
  \BibitemOpen
  \bibfield  {author} {\bibinfo {author} {\bibfnamefont {J.~W.}\ \bibnamefont {Maluf}},\ }\href {\doibase 10.1002/andp.201200272} {\bibfield  {journal} {\bibinfo  {journal} {Annalen der Physik}\ }\textbf {\bibinfo {volume} {525}},\ \bibinfo {pages} {339–357} (\bibinfo {year} {2013})}\BibitemShut {NoStop}%
\bibitem [{\citenamefont {Bahamonde}\ \emph {et~al.}(2023)\citenamefont {Bahamonde}, \citenamefont {Dialektopoulos} \emph {et~al.}}]{Bahamonde_2023}%
  \BibitemOpen
  \bibfield  {author} {\bibinfo {author} {\bibfnamefont {S.}~\bibnamefont {Bahamonde}}, \bibinfo {author} {\bibnamefont {Dialektopoulos}},  \emph {et~al.},\ }\href {\doibase 10.1088/1361-6633/ac9cef} {\bibfield  {journal} {\bibinfo  {journal} {Reports on Progress in Physics}\ }\textbf {\bibinfo {volume} {86}},\ \bibinfo {pages} {026901} (\bibinfo {year} {2023})}\BibitemShut {NoStop}%
\bibitem [{\citenamefont {Cai}\ \emph {et~al.}(2016)\citenamefont {Cai}, \citenamefont {Capozziello}, \citenamefont {De~Laurentis},\ and\ \citenamefont {Saridakis}}]{Cai_2016}%
  \BibitemOpen
  \bibfield  {author} {\bibinfo {author} {\bibfnamefont {Y.-F.}\ \bibnamefont {Cai}}, \bibinfo {author} {\bibfnamefont {S.}~\bibnamefont {Capozziello}}, \bibinfo {author} {\bibfnamefont {M.}~\bibnamefont {De~Laurentis}}, \ and\ \bibinfo {author} {\bibfnamefont {E.~N.}\ \bibnamefont {Saridakis}},\ }\href {\doibase 10.1088/0034-4885/79/10/106901} {\bibfield  {journal} {\bibinfo  {journal} {Reports on Progress in Physics}\ }\textbf {\bibinfo {volume} {79}},\ \bibinfo {pages} {106901} (\bibinfo {year} {2016})}\BibitemShut {NoStop}%
\bibitem [{\citenamefont {Krššák}\ \emph {et~al.}(2019)\citenamefont {Krššák}, \citenamefont {van~den Hoogen}, \citenamefont {Pereira}, \citenamefont {Böhmer},\ and\ \citenamefont {Coley}}]{Kr_k_2019}%
  \BibitemOpen
  \bibfield  {author} {\bibinfo {author} {\bibfnamefont {M.}~\bibnamefont {Krššák}}, \bibinfo {author} {\bibfnamefont {R.~J.}\ \bibnamefont {van~den Hoogen}}, \bibinfo {author} {\bibfnamefont {J.~G.}\ \bibnamefont {Pereira}}, \bibinfo {author} {\bibfnamefont {C.~G.}\ \bibnamefont {Böhmer}}, \ and\ \bibinfo {author} {\bibfnamefont {A.~A.}\ \bibnamefont {Coley}},\ }\href {\doibase 10.1088/1361-6382/ab2e1f} {\bibfield  {journal} {\bibinfo  {journal} {Classical and Quantum Gravity}\ }\textbf {\bibinfo {volume} {36}},\ \bibinfo {pages} {183001} (\bibinfo {year} {2019})}\BibitemShut {NoStop}%
\bibitem [{\citenamefont {Abdalla}\ \emph {et~al.}(2022)\citenamefont {Abdalla}, \citenamefont {Abellán} \emph {et~al.}}]{Abdalla_2022}%
  \BibitemOpen
  \bibfield  {author} {\bibinfo {author} {\bibfnamefont {E.}~\bibnamefont {Abdalla}}, \bibinfo {author} {\bibfnamefont {G.~F.}\ \bibnamefont {Abellán}},  \emph {et~al.},\ }\href {\doibase 10.1016/j.jheap.2022.04.002} {\bibfield  {journal} {\bibinfo  {journal} {Journal of High Energy Astrophysics}\ }\textbf {\bibinfo {volume} {34}},\ \bibinfo {pages} {49–211} (\bibinfo {year} {2022})}\BibitemShut {NoStop}%
\bibitem [{\citenamefont {Aghanim}\ \emph {et~al.}(2020)\citenamefont {Aghanim} \emph {et~al.}}]{Planck:2018vyg}%
  \BibitemOpen
  \bibfield  {author} {\bibinfo {author} {\bibfnamefont {N.}~\bibnamefont {Aghanim}} \emph {et~al.} (\bibinfo {collaboration} {Planck}),\ }\href {\doibase 10.1051/0004-6361/201833910} {\bibfield  {journal} {\bibinfo  {journal} {Astron. Astrophys.}\ }\textbf {\bibinfo {volume} {641}},\ \bibinfo {pages} {A6} (\bibinfo {year} {2020})},\ \bibinfo {note} {[Erratum: Astron.Astrophys. 652, C4 (2021)]},\ \Eprint {http://arxiv.org/abs/1807.06209} {arXiv:1807.06209 [astro-ph.CO]} \BibitemShut {NoStop}%
\bibitem [{\citenamefont {Riess}\ \emph {et~al.}(2022)\citenamefont {Riess} \emph {et~al.}}]{Riess:2021jrx}%
  \BibitemOpen
  \bibfield  {author} {\bibinfo {author} {\bibfnamefont {A.~G.}\ \bibnamefont {Riess}} \emph {et~al.},\ }\href {\doibase 10.3847/2041-8213/ac5c5b} {\bibfield  {journal} {\bibinfo  {journal} {Astrophys. J. Lett.}\ }\textbf {\bibinfo {volume} {934}},\ \bibinfo {pages} {L7} (\bibinfo {year} {2022})},\ \Eprint {http://arxiv.org/abs/2112.04510} {arXiv:2112.04510 [astro-ph.CO]} \BibitemShut {NoStop}%
\bibitem [{\citenamefont {Di~Valentino}\ \emph {et~al.}(2021{\natexlab{a}})\citenamefont {Di~Valentino}, \citenamefont {Mena}, \citenamefont {Pan}, \citenamefont {Visinelli}, \citenamefont {Yang}, \citenamefont {Melchiorri}, \citenamefont {Mota}, \citenamefont {Riess},\ and\ \citenamefont {Silk}}]{Di_Valentino_2021}%
  \BibitemOpen
  \bibfield  {author} {\bibinfo {author} {\bibfnamefont {E.}~\bibnamefont {Di~Valentino}}, \bibinfo {author} {\bibfnamefont {O.}~\bibnamefont {Mena}}, \bibinfo {author} {\bibfnamefont {S.}~\bibnamefont {Pan}}, \bibinfo {author} {\bibfnamefont {L.}~\bibnamefont {Visinelli}}, \bibinfo {author} {\bibfnamefont {W.}~\bibnamefont {Yang}}, \bibinfo {author} {\bibfnamefont {A.}~\bibnamefont {Melchiorri}}, \bibinfo {author} {\bibfnamefont {D.~F.}\ \bibnamefont {Mota}}, \bibinfo {author} {\bibfnamefont {A.~G.}\ \bibnamefont {Riess}}, \ and\ \bibinfo {author} {\bibfnamefont {J.}~\bibnamefont {Silk}},\ }\href {\doibase 10.1088/1361-6382/ac086d} {\bibfield  {journal} {\bibinfo  {journal} {Classical and Quantum Gravity}\ }\textbf {\bibinfo {volume} {38}},\ \bibinfo {pages} {153001} (\bibinfo {year} {2021}{\natexlab{a}})}\BibitemShut {NoStop}%
\bibitem [{\citenamefont {Perivolaropoulos}\ and\ \citenamefont {Skara}(2022)}]{Perivolaropoulos_2022}%
  \BibitemOpen
  \bibfield  {author} {\bibinfo {author} {\bibfnamefont {L.}~\bibnamefont {Perivolaropoulos}}\ and\ \bibinfo {author} {\bibfnamefont {F.}~\bibnamefont {Skara}},\ }\href {\doibase 10.1016/j.newar.2022.101659} {\bibfield  {journal} {\bibinfo  {journal} {New Astronomy Reviews}\ }\textbf {\bibinfo {volume} {95}},\ \bibinfo {pages} {101659} (\bibinfo {year} {2022})}\BibitemShut {NoStop}%
\bibitem [{\citenamefont {Dalal}\ \emph {et~al.}(2023)\citenamefont {Dalal}, \citenamefont {Li} \emph {et~al.}}]{Dalal_2023}%
  \BibitemOpen
  \bibfield  {author} {\bibinfo {author} {\bibfnamefont {R.}~\bibnamefont {Dalal}}, \bibinfo {author} {\bibfnamefont {X.}~\bibnamefont {Li}},  \emph {et~al.},\ }\href {\doibase 10.1103/physrevd.108.123519} {\bibfield  {journal} {\bibinfo  {journal} {Physical Review D}\ }\textbf {\bibinfo {volume} {108}} (\bibinfo {year} {2023}),\ 10.1103/physrevd.108.123519}\BibitemShut {NoStop}%
\bibitem [{\citenamefont {Asgari}\ \emph {et~al.}(2021)\citenamefont {Asgari} \emph {et~al.}}]{KiDS:2020suj}%
  \BibitemOpen
  \bibfield  {author} {\bibinfo {author} {\bibfnamefont {M.}~\bibnamefont {Asgari}} \emph {et~al.} (\bibinfo {collaboration} {KiDS}),\ }\href {\doibase 10.1051/0004-6361/202039070} {\bibfield  {journal} {\bibinfo  {journal} {Astron. Astrophys.}\ }\textbf {\bibinfo {volume} {645}},\ \bibinfo {pages} {A104} (\bibinfo {year} {2021})},\ \Eprint {http://arxiv.org/abs/2007.15633} {arXiv:2007.15633 [astro-ph.CO]} \BibitemShut {NoStop}%
\bibitem [{\citenamefont {Amon}\ \emph {et~al.}(2022)\citenamefont {Amon}, \citenamefont {Gruen}, \citenamefont {Troxel} \emph {et~al.}}]{Amon_2022}%
  \BibitemOpen
  \bibfield  {author} {\bibinfo {author} {\bibfnamefont {A.}~\bibnamefont {Amon}}, \bibinfo {author} {\bibfnamefont {D.}~\bibnamefont {Gruen}}, \bibinfo {author} {\bibnamefont {Troxel}},  \emph {et~al.},\ }\href {\doibase 10.1103/physrevd.105.023514} {\bibfield  {journal} {\bibinfo  {journal} {Physical Review D}\ }\textbf {\bibinfo {volume} {105}} (\bibinfo {year} {2022}),\ 10.1103/physrevd.105.023514}\BibitemShut {NoStop}%
\bibitem [{\citenamefont {Nunes}\ and\ \citenamefont {Vagnozzi}(2021)}]{Nunes_2021}%
  \BibitemOpen
  \bibfield  {author} {\bibinfo {author} {\bibfnamefont {R.~C.}\ \bibnamefont {Nunes}}\ and\ \bibinfo {author} {\bibfnamefont {S.}~\bibnamefont {Vagnozzi}},\ }\href {\doibase 10.1093/mnras/stab1613} {\bibfield  {journal} {\bibinfo  {journal} {Monthly Notices of the Royal Astronomical Society}\ }\textbf {\bibinfo {volume} {505}},\ \bibinfo {pages} {5427–5437} (\bibinfo {year} {2021})}\BibitemShut {NoStop}%
\bibitem [{\citenamefont {Skara}\ and\ \citenamefont {Perivolaropoulos}(2020)}]{Skara_2020}%
  \BibitemOpen
  \bibfield  {author} {\bibinfo {author} {\bibfnamefont {F.}~\bibnamefont {Skara}}\ and\ \bibinfo {author} {\bibfnamefont {L.}~\bibnamefont {Perivolaropoulos}},\ }\href {\doibase 10.1103/physrevd.101.063521} {\bibfield  {journal} {\bibinfo  {journal} {Physical Review D}\ }\textbf {\bibinfo {volume} {101}} (\bibinfo {year} {2020}),\ 10.1103/physrevd.101.063521}\BibitemShut {NoStop}%
\bibitem [{\citenamefont {Briffa}\ \emph {et~al.}(2022)\citenamefont {Briffa}, \citenamefont {Escamilla-Rivera}, \citenamefont {Said}, \citenamefont {Mifsud},\ and\ \citenamefont {Pullicino}}]{Briffa_2022}%
  \BibitemOpen
  \bibfield  {author} {\bibinfo {author} {\bibfnamefont {R.}~\bibnamefont {Briffa}}, \bibinfo {author} {\bibfnamefont {C.}~\bibnamefont {Escamilla-Rivera}}, \bibinfo {author} {\bibfnamefont {J.~L.}\ \bibnamefont {Said}}, \bibinfo {author} {\bibfnamefont {J.}~\bibnamefont {Mifsud}}, \ and\ \bibinfo {author} {\bibfnamefont {N.~L.}\ \bibnamefont {Pullicino}},\ }\href {\doibase 10.1140/epjp/s13360-022-02725-4} {\bibfield  {journal} {\bibinfo  {journal} {The European Physical Journal Plus}\ }\textbf {\bibinfo {volume} {137}} (\bibinfo {year} {2022}),\ 10.1140/epjp/s13360-022-02725-4}\BibitemShut {NoStop}%
\bibitem [{\citenamefont {Briffa}\ \emph {et~al.}(2023)\citenamefont {Briffa}, \citenamefont {Escamilla-Rivera}, \citenamefont {Said},\ and\ \citenamefont {Mifsud}}]{Briffa_2023}%
  \BibitemOpen
  \bibfield  {author} {\bibinfo {author} {\bibfnamefont {R.}~\bibnamefont {Briffa}}, \bibinfo {author} {\bibfnamefont {C.}~\bibnamefont {Escamilla-Rivera}}, \bibinfo {author} {\bibfnamefont {J.~L.}\ \bibnamefont {Said}}, \ and\ \bibinfo {author} {\bibfnamefont {J.}~\bibnamefont {Mifsud}},\ }\href {\doibase 10.1093/mnras/stad1384} {\bibfield  {journal} {\bibinfo  {journal} {Monthly Notices of the Royal Astronomical Society}\ }\textbf {\bibinfo {volume} {522}},\ \bibinfo {pages} {6024–6034} (\bibinfo {year} {2023})}\BibitemShut {NoStop}%
\bibitem [{\citenamefont {Sandoval-Orozco}\ \emph {et~al.}(2024{\natexlab{a}})\citenamefont {Sandoval-Orozco}, \citenamefont {Escamilla-Rivera}, \citenamefont {Briffa},\ and\ \citenamefont {Said}}]{sandovalorozco2024testingftcosmologieshii}%
  \BibitemOpen
  \bibfield  {author} {\bibinfo {author} {\bibfnamefont {R.}~\bibnamefont {Sandoval-Orozco}}, \bibinfo {author} {\bibfnamefont {C.}~\bibnamefont {Escamilla-Rivera}}, \bibinfo {author} {\bibfnamefont {R.}~\bibnamefont {Briffa}}, \ and\ \bibinfo {author} {\bibfnamefont {J.~L.}\ \bibnamefont {Said}},\ }\href {https://arxiv.org/abs/2405.06633} {\enquote {\bibinfo {title} {Testing $f(t)$ cosmologies with hii hubble diagram and cmb distance priors},}\ } (\bibinfo {year} {2024}{\natexlab{a}}),\ \Eprint {http://arxiv.org/abs/2405.06633} {arXiv:2405.06633 [astro-ph.CO]} \BibitemShut {NoStop}%
\bibitem [{\citenamefont {Zhadyranova}\ \emph {et~al.}(2024)\citenamefont {Zhadyranova}, \citenamefont {Koussour}, \citenamefont {Bekkhozhayev}, \citenamefont {Zhumabekova},\ and\ \citenamefont {Rayimbaev}}]{Zhadyranova_2024}%
  \BibitemOpen
  \bibfield  {author} {\bibinfo {author} {\bibfnamefont {A.}~\bibnamefont {Zhadyranova}}, \bibinfo {author} {\bibfnamefont {M.}~\bibnamefont {Koussour}}, \bibinfo {author} {\bibfnamefont {S.}~\bibnamefont {Bekkhozhayev}}, \bibinfo {author} {\bibfnamefont {V.}~\bibnamefont {Zhumabekova}}, \ and\ \bibinfo {author} {\bibfnamefont {J.}~\bibnamefont {Rayimbaev}},\ }\href {\doibase 10.1016/j.dark.2024.101514} {\bibfield  {journal} {\bibinfo  {journal} {Physics of the Dark Universe}\ }\textbf {\bibinfo {volume} {45}},\ \bibinfo {pages} {101514} (\bibinfo {year} {2024})}\BibitemShut {NoStop}%
\bibitem [{\citenamefont {Capozziello}\ \emph {et~al.}(2017)\citenamefont {Capozziello}, \citenamefont {D’Agostino},\ and\ \citenamefont {Luongo}}]{Capozziello_2017}%
  \BibitemOpen
  \bibfield  {author} {\bibinfo {author} {\bibfnamefont {S.}~\bibnamefont {Capozziello}}, \bibinfo {author} {\bibfnamefont {R.}~\bibnamefont {D’Agostino}}, \ and\ \bibinfo {author} {\bibfnamefont {O.}~\bibnamefont {Luongo}},\ }\href {\doibase 10.1007/s10714-017-2304-x} {\bibfield  {journal} {\bibinfo  {journal} {General Relativity and Gravitation}\ }\textbf {\bibinfo {volume} {49}} (\bibinfo {year} {2017}),\ 10.1007/s10714-017-2304-x}\BibitemShut {NoStop}%
\bibitem [{\citenamefont {Qi}\ \emph {et~al.}(2017)\citenamefont {Qi}, \citenamefont {Cao}, \citenamefont {Biesiada}, \citenamefont {Zheng},\ and\ \citenamefont {Zhu}}]{Qi_2017}%
  \BibitemOpen
  \bibfield  {author} {\bibinfo {author} {\bibfnamefont {J.-Z.}\ \bibnamefont {Qi}}, \bibinfo {author} {\bibfnamefont {S.}~\bibnamefont {Cao}}, \bibinfo {author} {\bibfnamefont {M.}~\bibnamefont {Biesiada}}, \bibinfo {author} {\bibfnamefont {X.}~\bibnamefont {Zheng}}, \ and\ \bibinfo {author} {\bibfnamefont {Z.-H.}\ \bibnamefont {Zhu}},\ }\href {\doibase 10.1140/epjc/s10052-017-5069-1} {\bibfield  {journal} {\bibinfo  {journal} {The European Physical Journal C}\ }\textbf {\bibinfo {volume} {77}} (\bibinfo {year} {2017}),\ 10.1140/epjc/s10052-017-5069-1}\BibitemShut {NoStop}%
\bibitem [{\citenamefont {Basilakos}\ \emph {et~al.}(2018)\citenamefont {Basilakos}, \citenamefont {Nesseris}, \citenamefont {Anagnostopoulos},\ and\ \citenamefont {Saridakis}}]{Basilakos_2018}%
  \BibitemOpen
  \bibfield  {author} {\bibinfo {author} {\bibfnamefont {S.}~\bibnamefont {Basilakos}}, \bibinfo {author} {\bibfnamefont {S.}~\bibnamefont {Nesseris}}, \bibinfo {author} {\bibfnamefont {F.~K.}\ \bibnamefont {Anagnostopoulos}}, \ and\ \bibinfo {author} {\bibfnamefont {E.~N.}\ \bibnamefont {Saridakis}},\ }\href {\doibase 10.1088/1475-7516/2018/08/008} {\bibfield  {journal} {\bibinfo  {journal} {Journal of Cosmology and Astroparticle Physics}\ }\textbf {\bibinfo {volume} {2018}},\ \bibinfo {pages} {008–008} (\bibinfo {year} {2018})}\BibitemShut {NoStop}%
\bibitem [{\citenamefont {El-Zant}\ \emph {et~al.}(2019)\citenamefont {El-Zant}, \citenamefont {El~Hanafy},\ and\ \citenamefont {Elgammal}}]{El_Zant_2019}%
  \BibitemOpen
  \bibfield  {author} {\bibinfo {author} {\bibfnamefont {A.}~\bibnamefont {El-Zant}}, \bibinfo {author} {\bibfnamefont {W.}~\bibnamefont {El~Hanafy}}, \ and\ \bibinfo {author} {\bibfnamefont {S.}~\bibnamefont {Elgammal}},\ }\href {\doibase 10.3847/1538-4357/aafa12} {\bibfield  {journal} {\bibinfo  {journal} {The Astrophysical Journal}\ }\textbf {\bibinfo {volume} {871}},\ \bibinfo {pages} {210} (\bibinfo {year} {2019})}\BibitemShut {NoStop}%
\bibitem [{\citenamefont {Said}\ \emph {et~al.}(2020)\citenamefont {Said}, \citenamefont {Mifsud}, \citenamefont {Parkinson}, \citenamefont {Saridakis}, \citenamefont {Sultana},\ and\ \citenamefont {Adami}}]{Said_2020}%
  \BibitemOpen
  \bibfield  {author} {\bibinfo {author} {\bibfnamefont {J.~L.}\ \bibnamefont {Said}}, \bibinfo {author} {\bibfnamefont {J.}~\bibnamefont {Mifsud}}, \bibinfo {author} {\bibfnamefont {D.}~\bibnamefont {Parkinson}}, \bibinfo {author} {\bibfnamefont {E.~N.}\ \bibnamefont {Saridakis}}, \bibinfo {author} {\bibfnamefont {J.}~\bibnamefont {Sultana}}, \ and\ \bibinfo {author} {\bibfnamefont {K.~Z.}\ \bibnamefont {Adami}},\ }\href {\doibase 10.1088/1475-7516/2020/11/047} {\bibfield  {journal} {\bibinfo  {journal} {Journal of Cosmology and Astroparticle Physics}\ }\textbf {\bibinfo {volume} {2020}},\ \bibinfo {pages} {047–047} (\bibinfo {year} {2020})}\BibitemShut {NoStop}%
\bibitem [{\citenamefont {Benetti}\ \emph {et~al.}(2020)\citenamefont {Benetti}, \citenamefont {Capozziello},\ and\ \citenamefont {Lambiase}}]{Benetti_2020}%
  \BibitemOpen
  \bibfield  {author} {\bibinfo {author} {\bibfnamefont {M.}~\bibnamefont {Benetti}}, \bibinfo {author} {\bibfnamefont {S.}~\bibnamefont {Capozziello}}, \ and\ \bibinfo {author} {\bibfnamefont {G.}~\bibnamefont {Lambiase}},\ }\href {\doibase 10.1093/mnras/staa3368} {\bibfield  {journal} {\bibinfo  {journal} {Monthly Notices of the Royal Astronomical Society}\ }\textbf {\bibinfo {volume} {500}},\ \bibinfo {pages} {1795–1805} (\bibinfo {year} {2020})}\BibitemShut {NoStop}%
\bibitem [{\citenamefont {dos Santos}\ \emph {et~al.}(2022)\citenamefont {dos Santos}, \citenamefont {Gonzalez},\ and\ \citenamefont {Silva}}]{dos_Santos_2022}%
  \BibitemOpen
  \bibfield  {author} {\bibinfo {author} {\bibfnamefont {F.~B.~M.}\ \bibnamefont {dos Santos}}, \bibinfo {author} {\bibfnamefont {J.~E.}\ \bibnamefont {Gonzalez}}, \ and\ \bibinfo {author} {\bibfnamefont {R.}~\bibnamefont {Silva}},\ }\href {\doibase 10.1140/epjc/s10052-022-10784-1} {\bibfield  {journal} {\bibinfo  {journal} {The European Physical Journal C}\ }\textbf {\bibinfo {volume} {82}} (\bibinfo {year} {2022}),\ 10.1140/epjc/s10052-022-10784-1}\BibitemShut {NoStop}%
\bibitem [{\citenamefont {Aljaf}\ \emph {et~al.}(2022)\citenamefont {Aljaf}, \citenamefont {Elizalde}, \citenamefont {Khurshudyan}, \citenamefont {Myrzakulov},\ and\ \citenamefont {Zhadyranova}}]{Aljaf_2022}%
  \BibitemOpen
  \bibfield  {author} {\bibinfo {author} {\bibfnamefont {M.}~\bibnamefont {Aljaf}}, \bibinfo {author} {\bibfnamefont {E.}~\bibnamefont {Elizalde}}, \bibinfo {author} {\bibfnamefont {M.}~\bibnamefont {Khurshudyan}}, \bibinfo {author} {\bibfnamefont {K.}~\bibnamefont {Myrzakulov}}, \ and\ \bibinfo {author} {\bibfnamefont {A.}~\bibnamefont {Zhadyranova}},\ }\href {\doibase 10.1140/epjc/s10052-022-11109-y} {\bibfield  {journal} {\bibinfo  {journal} {The European Physical Journal C}\ }\textbf {\bibinfo {volume} {82}} (\bibinfo {year} {2022}),\ 10.1140/epjc/s10052-022-11109-y}\BibitemShut {NoStop}%
\bibitem [{\citenamefont {Sabiee}\ \emph {et~al.}(2022)\citenamefont {Sabiee}, \citenamefont {Malekjani},\ and\ \citenamefont {Mohammad Zadeh~Jassur}}]{Sabiee_2022}%
  \BibitemOpen
  \bibfield  {author} {\bibinfo {author} {\bibfnamefont {M.}~\bibnamefont {Sabiee}}, \bibinfo {author} {\bibfnamefont {M.}~\bibnamefont {Malekjani}}, \ and\ \bibinfo {author} {\bibfnamefont {D.}~\bibnamefont {Mohammad Zadeh~Jassur}},\ }\href {\doibase 10.1093/mnras/stac2367} {\bibfield  {journal} {\bibinfo  {journal} {Monthly Notices of the Royal Astronomical Society}\ }\textbf {\bibinfo {volume} {516}},\ \bibinfo {pages} {2597–2613} (\bibinfo {year} {2022})}\BibitemShut {NoStop}%
\bibitem [{\citenamefont {dos Santos}(2023)}]{dos_Santos_2023}%
  \BibitemOpen
  \bibfield  {author} {\bibinfo {author} {\bibfnamefont {F.}~\bibnamefont {dos Santos}},\ }\href {\doibase 10.1088/1475-7516/2023/06/039} {\bibfield  {journal} {\bibinfo  {journal} {Journal of Cosmology and Astroparticle Physics}\ }\textbf {\bibinfo {volume} {2023}},\ \bibinfo {pages} {039} (\bibinfo {year} {2023})}\BibitemShut {NoStop}%
\bibitem [{\citenamefont {Kavya}\ \emph {et~al.}(2024)\citenamefont {Kavya}, \citenamefont {Mishra}, \citenamefont {Sahoo},\ and\ \citenamefont {Venkatesha}}]{Kavya_2024}%
  \BibitemOpen
  \bibfield  {author} {\bibinfo {author} {\bibfnamefont {N.~S.}\ \bibnamefont {Kavya}}, \bibinfo {author} {\bibfnamefont {S.~S.}\ \bibnamefont {Mishra}}, \bibinfo {author} {\bibfnamefont {P.~K.}\ \bibnamefont {Sahoo}}, \ and\ \bibinfo {author} {\bibfnamefont {V.}~\bibnamefont {Venkatesha}},\ }\href {\doibase 10.1093/mnras/stae1723} {\bibfield  {journal} {\bibinfo  {journal} {Monthly Notices of the Royal Astronomical Society}\ }\textbf {\bibinfo {volume} {532}},\ \bibinfo {pages} {3126–3133} (\bibinfo {year} {2024})}\BibitemShut {NoStop}%
\bibitem [{\citenamefont {Nunes}\ \emph {et~al.}(2016)\citenamefont {Nunes}, \citenamefont {Pan},\ and\ \citenamefont {Saridakis}}]{Nunes_2016}%
  \BibitemOpen
  \bibfield  {author} {\bibinfo {author} {\bibfnamefont {R.~C.}\ \bibnamefont {Nunes}}, \bibinfo {author} {\bibfnamefont {S.}~\bibnamefont {Pan}}, \ and\ \bibinfo {author} {\bibfnamefont {E.~N.}\ \bibnamefont {Saridakis}},\ }\href {\doibase 10.1088/1475-7516/2016/08/011} {\bibfield  {journal} {\bibinfo  {journal} {Journal of Cosmology and Astroparticle Physics}\ }\textbf {\bibinfo {volume} {2016}},\ \bibinfo {pages} {011–011} (\bibinfo {year} {2016})}\BibitemShut {NoStop}%
\bibitem [{\citenamefont {Capozziello}\ \emph {et~al.}(2024)\citenamefont {Capozziello}, \citenamefont {Caruana}, \citenamefont {Farrugia}, \citenamefont {Levi~Said},\ and\ \citenamefont {Sultana}}]{Capozziello_2024}%
  \BibitemOpen
  \bibfield  {author} {\bibinfo {author} {\bibfnamefont {S.}~\bibnamefont {Capozziello}}, \bibinfo {author} {\bibfnamefont {M.}~\bibnamefont {Caruana}}, \bibinfo {author} {\bibfnamefont {G.}~\bibnamefont {Farrugia}}, \bibinfo {author} {\bibfnamefont {J.}~\bibnamefont {Levi~Said}}, \ and\ \bibinfo {author} {\bibfnamefont {J.}~\bibnamefont {Sultana}},\ }\href {\doibase 10.1007/s10714-024-03204-0} {\bibfield  {journal} {\bibinfo  {journal} {General Relativity and Gravitation}\ }\textbf {\bibinfo {volume} {56}} (\bibinfo {year} {2024}),\ 10.1007/s10714-024-03204-0}\BibitemShut {NoStop}%
\bibitem [{\citenamefont {Aguilar}\ \emph {et~al.}(2024)\citenamefont {Aguilar}, \citenamefont {Escamilla-Rivera}, \citenamefont {Said},\ and\ \citenamefont {Mifsud}}]{aguilar2024nonfluidlikeboltzmanncode}%
  \BibitemOpen
  \bibfield  {author} {\bibinfo {author} {\bibfnamefont {A.}~\bibnamefont {Aguilar}}, \bibinfo {author} {\bibfnamefont {C.}~\bibnamefont {Escamilla-Rivera}}, \bibinfo {author} {\bibfnamefont {J.~L.}\ \bibnamefont {Said}}, \ and\ \bibinfo {author} {\bibfnamefont {J.}~\bibnamefont {Mifsud}},\ }\href {https://arxiv.org/abs/2403.13708} {\enquote {\bibinfo {title} {Non-fluid like boltzmann code architecture for early times $f(t)$ cosmologies},}\ } (\bibinfo {year} {2024}),\ \Eprint {http://arxiv.org/abs/2403.13708} {arXiv:2403.13708 [gr-qc]} \BibitemShut {NoStop}%
\bibitem [{\citenamefont {Briffa}\ \emph {et~al.}(2024)\citenamefont {Briffa}, \citenamefont {Escamilla-Rivera}, \citenamefont {Levi~Said},\ and\ \citenamefont {Mifsud}}]{Briffa_2024}%
  \BibitemOpen
  \bibfield  {author} {\bibinfo {author} {\bibfnamefont {R.}~\bibnamefont {Briffa}}, \bibinfo {author} {\bibfnamefont {C.}~\bibnamefont {Escamilla-Rivera}}, \bibinfo {author} {\bibfnamefont {J.}~\bibnamefont {Levi~Said}}, \ and\ \bibinfo {author} {\bibfnamefont {J.}~\bibnamefont {Mifsud}},\ }\href {\doibase 10.1093/mnras/stae103} {\bibfield  {journal} {\bibinfo  {journal} {Monthly Notices of the Royal Astronomical Society}\ }\textbf {\bibinfo {volume} {528}},\ \bibinfo {pages} {2711–2727} (\bibinfo {year} {2024})}\BibitemShut {NoStop}%
\bibitem [{\citenamefont {Anagnostopoulos}\ \emph {et~al.}(2019)\citenamefont {Anagnostopoulos}, \citenamefont {Basilakos},\ and\ \citenamefont {Saridakis}}]{Anagnostopoulos_2019}%
  \BibitemOpen
  \bibfield  {author} {\bibinfo {author} {\bibfnamefont {F.~K.}\ \bibnamefont {Anagnostopoulos}}, \bibinfo {author} {\bibfnamefont {S.}~\bibnamefont {Basilakos}}, \ and\ \bibinfo {author} {\bibfnamefont {E.~N.}\ \bibnamefont {Saridakis}},\ }\href {\doibase 10.1103/physrevd.100.083517} {\bibfield  {journal} {\bibinfo  {journal} {Physical Review D}\ }\textbf {\bibinfo {volume} {100}} (\bibinfo {year} {2019}),\ 10.1103/physrevd.100.083517}\BibitemShut {NoStop}%
\bibitem [{\citenamefont {Sandoval-Orozco}\ \emph {et~al.}(2024{\natexlab{b}})\citenamefont {Sandoval-Orozco}, \citenamefont {Escamilla-Rivera}, \citenamefont {Briffa},\ and\ \citenamefont {Levi~Said}}]{Sandoval_Orozco_2024}%
  \BibitemOpen
  \bibfield  {author} {\bibinfo {author} {\bibfnamefont {R.}~\bibnamefont {Sandoval-Orozco}}, \bibinfo {author} {\bibfnamefont {C.}~\bibnamefont {Escamilla-Rivera}}, \bibinfo {author} {\bibfnamefont {R.}~\bibnamefont {Briffa}}, \ and\ \bibinfo {author} {\bibfnamefont {J.}~\bibnamefont {Levi~Said}},\ }\href {\doibase 10.1016/j.dark.2023.101407} {\bibfield  {journal} {\bibinfo  {journal} {Physics of the Dark Universe}\ }\textbf {\bibinfo {volume} {43}},\ \bibinfo {pages} {101407} (\bibinfo {year} {2024}{\natexlab{b}})}\BibitemShut {NoStop}%
\bibitem [{\citenamefont {Nunes}(2018)}]{Nunes_2018}%
  \BibitemOpen
  \bibfield  {author} {\bibinfo {author} {\bibfnamefont {R.~C.}\ \bibnamefont {Nunes}},\ }\href {\doibase 10.1088/1475-7516/2018/05/052} {\bibfield  {journal} {\bibinfo  {journal} {Journal of Cosmology and Astroparticle Physics}\ }\textbf {\bibinfo {volume} {2018}},\ \bibinfo {pages} {052–052} (\bibinfo {year} {2018})}\BibitemShut {NoStop}%
\bibitem [{\citenamefont {Kumar}\ \emph {et~al.}(2023)\citenamefont {Kumar}, \citenamefont {Nunes},\ and\ \citenamefont {Yadav}}]{Kumar_2023}%
  \BibitemOpen
  \bibfield  {author} {\bibinfo {author} {\bibfnamefont {S.}~\bibnamefont {Kumar}}, \bibinfo {author} {\bibfnamefont {R.~C.}\ \bibnamefont {Nunes}}, \ and\ \bibinfo {author} {\bibfnamefont {P.}~\bibnamefont {Yadav}},\ }\href {\doibase 10.1103/physrevd.107.063529} {\bibfield  {journal} {\bibinfo  {journal} {Physical Review D}\ }\textbf {\bibinfo {volume} {107}} (\bibinfo {year} {2023}),\ 10.1103/physrevd.107.063529}\BibitemShut {NoStop}%
\bibitem [{\citenamefont {Wang}\ and\ \citenamefont {Mota}(2020)}]{Wang_2020}%
  \BibitemOpen
  \bibfield  {author} {\bibinfo {author} {\bibfnamefont {D.}~\bibnamefont {Wang}}\ and\ \bibinfo {author} {\bibfnamefont {D.}~\bibnamefont {Mota}},\ }\href {\doibase 10.1103/physrevd.102.063530} {\bibfield  {journal} {\bibinfo  {journal} {Physical Review D}\ }\textbf {\bibinfo {volume} {102}} (\bibinfo {year} {2020}),\ 10.1103/physrevd.102.063530}\BibitemShut {NoStop}%
\bibitem [{\citenamefont {Bengochea}\ and\ \citenamefont {Ferraro}(2009)}]{Bengochea:2008gz}%
  \BibitemOpen
  \bibfield  {author} {\bibinfo {author} {\bibfnamefont {G.~R.}\ \bibnamefont {Bengochea}}\ and\ \bibinfo {author} {\bibfnamefont {R.}~\bibnamefont {Ferraro}},\ }\href {\doibase 10.1103/PhysRevD.79.124019} {\bibfield  {journal} {\bibinfo  {journal} {Phys. Rev. D}\ }\textbf {\bibinfo {volume} {79}},\ \bibinfo {pages} {124019} (\bibinfo {year} {2009})},\ \Eprint {http://arxiv.org/abs/0812.1205} {arXiv:0812.1205 [astro-ph]} \BibitemShut {NoStop}%
\bibitem [{\citenamefont {Linder}(2010)}]{Linder:2010py}%
  \BibitemOpen
  \bibfield  {author} {\bibinfo {author} {\bibfnamefont {E.~V.}\ \bibnamefont {Linder}},\ }\href {\doibase 10.1103/PhysRevD.81.127301} {\bibfield  {journal} {\bibinfo  {journal} {Phys. Rev. D}\ }\textbf {\bibinfo {volume} {81}},\ \bibinfo {pages} {127301} (\bibinfo {year} {2010})},\ \bibinfo {note} {[Erratum: Phys.Rev.D 82, 109902 (2010)]},\ \Eprint {http://arxiv.org/abs/1005.3039} {arXiv:1005.3039 [astro-ph.CO]} \BibitemShut {NoStop}%
\bibitem [{\citenamefont {Wu}\ and\ \citenamefont {Yu}(2011)}]{Wu:2010av}%
  \BibitemOpen
  \bibfield  {author} {\bibinfo {author} {\bibfnamefont {P.}~\bibnamefont {Wu}}\ and\ \bibinfo {author} {\bibfnamefont {H.~W.}\ \bibnamefont {Yu}},\ }\href {\doibase 10.1140/epjc/s10052-011-1552-2} {\bibfield  {journal} {\bibinfo  {journal} {Eur. Phys. J. C}\ }\textbf {\bibinfo {volume} {71}},\ \bibinfo {pages} {1552} (\bibinfo {year} {2011})},\ \Eprint {http://arxiv.org/abs/1008.3669} {arXiv:1008.3669 [gr-qc]} \BibitemShut {NoStop}%
\bibitem [{\citenamefont {Karami}\ and\ \citenamefont {Abdolmaleki}(2013)}]{Karami:2010bys}%
  \BibitemOpen
  \bibfield  {author} {\bibinfo {author} {\bibfnamefont {K.}~\bibnamefont {Karami}}\ and\ \bibinfo {author} {\bibfnamefont {A.}~\bibnamefont {Abdolmaleki}},\ }\href {\doibase 10.1088/1674-4527/13/7/001} {\bibfield  {journal} {\bibinfo  {journal} {Res. Astron. Astrophys.}\ }\textbf {\bibinfo {volume} {13}},\ \bibinfo {pages} {757} (\bibinfo {year} {2013})},\ \Eprint {http://arxiv.org/abs/1009.2459} {arXiv:1009.2459 [gr-qc]} \BibitemShut {NoStop}%
\bibitem [{\citenamefont {Bamba}\ \emph {et~al.}(2011)\citenamefont {Bamba}, \citenamefont {Geng}, \citenamefont {Lee},\ and\ \citenamefont {Luo}}]{Bamba:2010wb}%
  \BibitemOpen
  \bibfield  {author} {\bibinfo {author} {\bibfnamefont {K.}~\bibnamefont {Bamba}}, \bibinfo {author} {\bibfnamefont {C.-Q.}\ \bibnamefont {Geng}}, \bibinfo {author} {\bibfnamefont {C.-C.}\ \bibnamefont {Lee}}, \ and\ \bibinfo {author} {\bibfnamefont {L.-W.}\ \bibnamefont {Luo}},\ }\href {\doibase 10.1088/1475-7516/2011/01/021} {\bibfield  {journal} {\bibinfo  {journal} {JCAP}\ }\textbf {\bibinfo {volume} {01}},\ \bibinfo {pages} {021} (\bibinfo {year} {2011})},\ \Eprint {http://arxiv.org/abs/1011.0508} {arXiv:1011.0508 [astro-ph.CO]} \BibitemShut {NoStop}%
\bibitem [{\citenamefont {Cardone}\ \emph {et~al.}(2012)\citenamefont {Cardone}, \citenamefont {Radicella},\ and\ \citenamefont {Camera}}]{Cardone:2012xq}%
  \BibitemOpen
  \bibfield  {author} {\bibinfo {author} {\bibfnamefont {V.~F.}\ \bibnamefont {Cardone}}, \bibinfo {author} {\bibfnamefont {N.}~\bibnamefont {Radicella}}, \ and\ \bibinfo {author} {\bibfnamefont {S.}~\bibnamefont {Camera}},\ }\href {\doibase 10.1103/PhysRevD.85.124007} {\bibfield  {journal} {\bibinfo  {journal} {Phys. Rev. D}\ }\textbf {\bibinfo {volume} {85}},\ \bibinfo {pages} {124007} (\bibinfo {year} {2012})},\ \Eprint {http://arxiv.org/abs/1204.5294} {arXiv:1204.5294 [astro-ph.CO]} \BibitemShut {NoStop}%
\bibitem [{\citenamefont {Di~Valentino}\ \emph {et~al.}(2021{\natexlab{b}})\citenamefont {Di~Valentino}, \citenamefont {Mukherjee},\ and\ \citenamefont {Sen}}]{DiValentino:2020naf}%
  \BibitemOpen
  \bibfield  {author} {\bibinfo {author} {\bibfnamefont {E.}~\bibnamefont {Di~Valentino}}, \bibinfo {author} {\bibfnamefont {A.}~\bibnamefont {Mukherjee}}, \ and\ \bibinfo {author} {\bibfnamefont {A.~A.}\ \bibnamefont {Sen}},\ }\href {\doibase 10.3390/e23040404} {\bibfield  {journal} {\bibinfo  {journal} {Entropy}\ }\textbf {\bibinfo {volume} {23}},\ \bibinfo {pages} {404} (\bibinfo {year} {2021}{\natexlab{b}})},\ \Eprint {http://arxiv.org/abs/2005.12587} {arXiv:2005.12587 [astro-ph.CO]} \BibitemShut {NoStop}%
\bibitem [{\citenamefont {Adil}\ \emph {et~al.}(2024)\citenamefont {Adil}, \citenamefont {Akarsu}, \citenamefont {Di~Valentino}, \citenamefont {Nunes}, \citenamefont {\"Oz\"ulker}, \citenamefont {Sen},\ and\ \citenamefont {Specogna}}]{Adil:2023exv}%
  \BibitemOpen
  \bibfield  {author} {\bibinfo {author} {\bibfnamefont {S.~A.}\ \bibnamefont {Adil}}, \bibinfo {author} {\bibfnamefont {O.}~\bibnamefont {Akarsu}}, \bibinfo {author} {\bibfnamefont {E.}~\bibnamefont {Di~Valentino}}, \bibinfo {author} {\bibfnamefont {R.~C.}\ \bibnamefont {Nunes}}, \bibinfo {author} {\bibfnamefont {E.}~\bibnamefont {\"Oz\"ulker}}, \bibinfo {author} {\bibfnamefont {A.~A.}\ \bibnamefont {Sen}}, \ and\ \bibinfo {author} {\bibfnamefont {E.}~\bibnamefont {Specogna}},\ }\href {\doibase 10.1103/PhysRevD.109.023527} {\bibfield  {journal} {\bibinfo  {journal} {Phys. Rev. D}\ }\textbf {\bibinfo {volume} {109}},\ \bibinfo {pages} {023527} (\bibinfo {year} {2024})},\ \Eprint {http://arxiv.org/abs/2306.08046} {arXiv:2306.08046 [astro-ph.CO]} \BibitemShut {NoStop}%
\bibitem [{\citenamefont {Akarsu}\ \emph {et~al.}(2021)\citenamefont {Akarsu}, \citenamefont {Kumar}, \citenamefont {\"Oz\"ulker},\ and\ \citenamefont {Vazquez}}]{Akarsu:2021fol}%
  \BibitemOpen
  \bibfield  {author} {\bibinfo {author} {\bibfnamefont {O.}~\bibnamefont {Akarsu}}, \bibinfo {author} {\bibfnamefont {S.}~\bibnamefont {Kumar}}, \bibinfo {author} {\bibfnamefont {E.}~\bibnamefont {\"Oz\"ulker}}, \ and\ \bibinfo {author} {\bibfnamefont {J.~A.}\ \bibnamefont {Vazquez}},\ }\href {\doibase 10.1103/PhysRevD.104.123512} {\bibfield  {journal} {\bibinfo  {journal} {Phys. Rev. D}\ }\textbf {\bibinfo {volume} {104}},\ \bibinfo {pages} {123512} (\bibinfo {year} {2021})},\ \Eprint {http://arxiv.org/abs/2108.09239} {arXiv:2108.09239 [astro-ph.CO]} \BibitemShut {NoStop}%
\bibitem [{\citenamefont {Akarsu}\ \emph {et~al.}(2023{\natexlab{a}})\citenamefont {Akarsu}, \citenamefont {Kumar}, \citenamefont {\"Oz\"ulker}, \citenamefont {Vazquez},\ and\ \citenamefont {Yadav}}]{Akarsu:2022typ}%
  \BibitemOpen
  \bibfield  {author} {\bibinfo {author} {\bibfnamefont {O.}~\bibnamefont {Akarsu}}, \bibinfo {author} {\bibfnamefont {S.}~\bibnamefont {Kumar}}, \bibinfo {author} {\bibfnamefont {E.}~\bibnamefont {\"Oz\"ulker}}, \bibinfo {author} {\bibfnamefont {J.~A.}\ \bibnamefont {Vazquez}}, \ and\ \bibinfo {author} {\bibfnamefont {A.}~\bibnamefont {Yadav}},\ }\href {\doibase 10.1103/PhysRevD.108.023513} {\bibfield  {journal} {\bibinfo  {journal} {Phys. Rev. D}\ }\textbf {\bibinfo {volume} {108}},\ \bibinfo {pages} {023513} (\bibinfo {year} {2023}{\natexlab{a}})},\ \Eprint {http://arxiv.org/abs/2211.05742} {arXiv:2211.05742 [astro-ph.CO]} \BibitemShut {NoStop}%
\bibitem [{\citenamefont {Akarsu}\ \emph {et~al.}(2023{\natexlab{b}})\citenamefont {Akarsu}, \citenamefont {Di~Valentino}, \citenamefont {Kumar}, \citenamefont {Nunes}, \citenamefont {Vazquez},\ and\ \citenamefont {Yadav}}]{Akarsu:2023mfb}%
  \BibitemOpen
  \bibfield  {author} {\bibinfo {author} {\bibfnamefont {O.}~\bibnamefont {Akarsu}}, \bibinfo {author} {\bibfnamefont {E.}~\bibnamefont {Di~Valentino}}, \bibinfo {author} {\bibfnamefont {S.}~\bibnamefont {Kumar}}, \bibinfo {author} {\bibfnamefont {R.~C.}\ \bibnamefont {Nunes}}, \bibinfo {author} {\bibfnamefont {J.~A.}\ \bibnamefont {Vazquez}}, \ and\ \bibinfo {author} {\bibfnamefont {A.}~\bibnamefont {Yadav}},\ }\href@noop {} {\  (\bibinfo {year} {2023}{\natexlab{b}})},\ \Eprint {http://arxiv.org/abs/2307.10899} {arXiv:2307.10899 [astro-ph.CO]} \BibitemShut {NoStop}%
\bibitem [{\citenamefont {Akarsu}\ \emph {et~al.}(2020)\citenamefont {Akarsu}, \citenamefont {Barrow}, \citenamefont {Escamilla},\ and\ \citenamefont {Vazquez}}]{Akarsu:2019hmw}%
  \BibitemOpen
  \bibfield  {author} {\bibinfo {author} {\bibfnamefont {O.}~\bibnamefont {Akarsu}}, \bibinfo {author} {\bibfnamefont {J.~D.}\ \bibnamefont {Barrow}}, \bibinfo {author} {\bibfnamefont {L.~A.}\ \bibnamefont {Escamilla}}, \ and\ \bibinfo {author} {\bibfnamefont {J.~A.}\ \bibnamefont {Vazquez}},\ }\href {\doibase 10.1103/PhysRevD.101.063528} {\bibfield  {journal} {\bibinfo  {journal} {Phys. Rev. D}\ }\textbf {\bibinfo {volume} {101}},\ \bibinfo {pages} {063528} (\bibinfo {year} {2020})},\ \Eprint {http://arxiv.org/abs/1912.08751} {arXiv:1912.08751 [astro-ph.CO]} \BibitemShut {NoStop}%
\bibitem [{\citenamefont {Akarsu}\ \emph {et~al.}(2024{\natexlab{a}})\citenamefont {Akarsu}, \citenamefont {De~Felice}, \citenamefont {Di~Valentino}, \citenamefont {Kumar}, \citenamefont {Nunes}, \citenamefont {Ozulker}, \citenamefont {Vazquez},\ and\ \citenamefont {Yadav}}]{Akarsu:2024qsi}%
  \BibitemOpen
  \bibfield  {author} {\bibinfo {author} {\bibfnamefont {O.}~\bibnamefont {Akarsu}}, \bibinfo {author} {\bibfnamefont {A.}~\bibnamefont {De~Felice}}, \bibinfo {author} {\bibfnamefont {E.}~\bibnamefont {Di~Valentino}}, \bibinfo {author} {\bibfnamefont {S.}~\bibnamefont {Kumar}}, \bibinfo {author} {\bibfnamefont {R.~C.}\ \bibnamefont {Nunes}}, \bibinfo {author} {\bibfnamefont {E.}~\bibnamefont {Ozulker}}, \bibinfo {author} {\bibfnamefont {J.~A.}\ \bibnamefont {Vazquez}}, \ and\ \bibinfo {author} {\bibfnamefont {A.}~\bibnamefont {Yadav}},\ }\href@noop {} {\  (\bibinfo {year} {2024}{\natexlab{a}})},\ \Eprint {http://arxiv.org/abs/2402.07716} {arXiv:2402.07716 [astro-ph.CO]} \BibitemShut {NoStop}%
\bibitem [{\citenamefont {Akarsu}\ \emph {et~al.}(2024{\natexlab{b}})\citenamefont {Akarsu}, \citenamefont {De~Felice}, \citenamefont {Di~Valentino}, \citenamefont {Kumar}, \citenamefont {Nunes}, \citenamefont {Ozulker}, \citenamefont {Vazquez},\ and\ \citenamefont {Yadav}}]{Akarsu:2024eoo}%
  \BibitemOpen
  \bibfield  {author} {\bibinfo {author} {\bibfnamefont {O.}~\bibnamefont {Akarsu}}, \bibinfo {author} {\bibfnamefont {A.}~\bibnamefont {De~Felice}}, \bibinfo {author} {\bibfnamefont {E.}~\bibnamefont {Di~Valentino}}, \bibinfo {author} {\bibfnamefont {S.}~\bibnamefont {Kumar}}, \bibinfo {author} {\bibfnamefont {R.~C.}\ \bibnamefont {Nunes}}, \bibinfo {author} {\bibfnamefont {E.}~\bibnamefont {Ozulker}}, \bibinfo {author} {\bibfnamefont {J.~A.}\ \bibnamefont {Vazquez}}, \ and\ \bibinfo {author} {\bibfnamefont {A.}~\bibnamefont {Yadav}},\ }\href@noop {} {\  (\bibinfo {year} {2024}{\natexlab{b}})},\ \Eprint {http://arxiv.org/abs/2406.07526} {arXiv:2406.07526 [astro-ph.CO]} \BibitemShut {NoStop}%
\bibitem [{\citenamefont {Adame}\ \emph {et~al.}(2024)\citenamefont {Adame} \emph {et~al.}}]{DESI:2024mwx}%
  \BibitemOpen
  \bibfield  {author} {\bibinfo {author} {\bibfnamefont {A.~G.}\ \bibnamefont {Adame}} \emph {et~al.} (\bibinfo {collaboration} {DESI}),\ }\href@noop {} {\  (\bibinfo {year} {2024})},\ \Eprint {http://arxiv.org/abs/2404.03002} {arXiv:2404.03002 [astro-ph.CO]} \BibitemShut {NoStop}%
\bibitem [{\citenamefont {Calderon}\ \emph {et~al.}(2024)\citenamefont {Calderon} \emph {et~al.}}]{DESI:2024aqx}%
  \BibitemOpen
  \bibfield  {author} {\bibinfo {author} {\bibfnamefont {R.}~\bibnamefont {Calderon}} \emph {et~al.} (\bibinfo {collaboration} {DESI}),\ }\href {\doibase 10.1088/1475-7516/2024/10/048} {\bibfield  {journal} {\bibinfo  {journal} {JCAP}\ }\textbf {\bibinfo {volume} {10}},\ \bibinfo {pages} {048} (\bibinfo {year} {2024})},\ \Eprint {http://arxiv.org/abs/2405.04216} {arXiv:2405.04216 [astro-ph.CO]} \BibitemShut {NoStop}%
\bibitem [{\citenamefont {Escamilla}\ \emph {et~al.}(2024)\citenamefont {Escamilla}, \citenamefont {\"Oz\"ulker}, \citenamefont {Akarsu}, \citenamefont {Di~Valentino},\ and\ \citenamefont {V\'azquez}}]{Escamilla:2024ahl}%
  \BibitemOpen
  \bibfield  {author} {\bibinfo {author} {\bibfnamefont {L.~A.}\ \bibnamefont {Escamilla}}, \bibinfo {author} {\bibfnamefont {E.}~\bibnamefont {\"Oz\"ulker}}, \bibinfo {author} {\bibfnamefont {O.}~\bibnamefont {Akarsu}}, \bibinfo {author} {\bibfnamefont {E.}~\bibnamefont {Di~Valentino}}, \ and\ \bibinfo {author} {\bibfnamefont {J.~A.}\ \bibnamefont {V\'azquez}},\ }\href@noop {} {\  (\bibinfo {year} {2024})},\ \Eprint {http://arxiv.org/abs/2408.12516} {arXiv:2408.12516 [astro-ph.CO]} \BibitemShut {NoStop}%
\bibitem [{\citenamefont {Escamilla}\ \emph {et~al.}(2023)\citenamefont {Escamilla}, \citenamefont {Akarsu}, \citenamefont {Di~Valentino},\ and\ \citenamefont {Vazquez}}]{Escamilla:2023shf}%
  \BibitemOpen
  \bibfield  {author} {\bibinfo {author} {\bibfnamefont {L.~A.}\ \bibnamefont {Escamilla}}, \bibinfo {author} {\bibfnamefont {O.}~\bibnamefont {Akarsu}}, \bibinfo {author} {\bibfnamefont {E.}~\bibnamefont {Di~Valentino}}, \ and\ \bibinfo {author} {\bibfnamefont {J.~A.}\ \bibnamefont {Vazquez}},\ }\href {\doibase 10.1088/1475-7516/2023/11/051} {\bibfield  {journal} {\bibinfo  {journal} {JCAP}\ }\textbf {\bibinfo {volume} {11}},\ \bibinfo {pages} {051} (\bibinfo {year} {2023})},\ \Eprint {http://arxiv.org/abs/2305.16290} {arXiv:2305.16290 [astro-ph.CO]} \BibitemShut {NoStop}%
\bibitem [{\citenamefont {Sabogal}\ \emph {et~al.}(2024)\citenamefont {Sabogal}, \citenamefont {Akarsu}, \citenamefont {Bonilla}, \citenamefont {Di~Valentino},\ and\ \citenamefont {Nunes}}]{Sabogal:2024qxs}%
  \BibitemOpen
  \bibfield  {author} {\bibinfo {author} {\bibfnamefont {M.~A.}\ \bibnamefont {Sabogal}}, \bibinfo {author} {\bibfnamefont {O.}~\bibnamefont {Akarsu}}, \bibinfo {author} {\bibfnamefont {A.}~\bibnamefont {Bonilla}}, \bibinfo {author} {\bibfnamefont {E.}~\bibnamefont {Di~Valentino}}, \ and\ \bibinfo {author} {\bibfnamefont {R.~C.}\ \bibnamefont {Nunes}},\ }\href {\doibase 10.1140/epjc/s10052-024-13081-1} {\bibfield  {journal} {\bibinfo  {journal} {Eur. Phys. J. C}\ }\textbf {\bibinfo {volume} {84}},\ \bibinfo {pages} {703} (\bibinfo {year} {2024})},\ \Eprint {http://arxiv.org/abs/2407.04223} {arXiv:2407.04223 [astro-ph.CO]} \BibitemShut {NoStop}%
\bibitem [{\citenamefont {Anchordoqui}\ \emph {et~al.}(2024{\natexlab{a}})\citenamefont {Anchordoqui}, \citenamefont {Antoniadis},\ and\ \citenamefont {Lust}}]{Anchordoqui:2023woo}%
  \BibitemOpen
  \bibfield  {author} {\bibinfo {author} {\bibfnamefont {L.~A.}\ \bibnamefont {Anchordoqui}}, \bibinfo {author} {\bibfnamefont {I.}~\bibnamefont {Antoniadis}}, \ and\ \bibinfo {author} {\bibfnamefont {D.}~\bibnamefont {Lust}},\ }\href {\doibase 10.1016/j.physletb.2024.138775} {\bibfield  {journal} {\bibinfo  {journal} {Phys. Lett. B}\ }\textbf {\bibinfo {volume} {855}},\ \bibinfo {pages} {138775} (\bibinfo {year} {2024}{\natexlab{a}})},\ \Eprint {http://arxiv.org/abs/2312.12352} {arXiv:2312.12352 [hep-th]} \BibitemShut {NoStop}%
\bibitem [{\citenamefont {Anchordoqui}\ \emph {et~al.}(2024{\natexlab{b}})\citenamefont {Anchordoqui}, \citenamefont {Antoniadis}, \citenamefont {Lust}, \citenamefont {Noble},\ and\ \citenamefont {Soriano}}]{Anchordoqui:2024gfa}%
  \BibitemOpen
  \bibfield  {author} {\bibinfo {author} {\bibfnamefont {L.~A.}\ \bibnamefont {Anchordoqui}}, \bibinfo {author} {\bibfnamefont {I.}~\bibnamefont {Antoniadis}}, \bibinfo {author} {\bibfnamefont {D.}~\bibnamefont {Lust}}, \bibinfo {author} {\bibfnamefont {N.~T.}\ \bibnamefont {Noble}}, \ and\ \bibinfo {author} {\bibfnamefont {J.~F.}\ \bibnamefont {Soriano}},\ }\href {\doibase 10.1016/j.dark.2024.101715} {\bibfield  {journal} {\bibinfo  {journal} {Phys. Dark Univ.}\ }\textbf {\bibinfo {volume} {46}},\ \bibinfo {pages} {101715} (\bibinfo {year} {2024}{\natexlab{b}})},\ \Eprint {http://arxiv.org/abs/2404.17334} {arXiv:2404.17334 [astro-ph.CO]} \BibitemShut {NoStop}%
\bibitem [{\citenamefont {Anchordoqui}\ \emph {et~al.}(2024{\natexlab{c}})\citenamefont {Anchordoqui}, \citenamefont {Antoniadis}, \citenamefont {Bielli}, \citenamefont {Chatrabhuti},\ and\ \citenamefont {Isono}}]{Anchordoqui:2024dqc}%
  \BibitemOpen
  \bibfield  {author} {\bibinfo {author} {\bibfnamefont {L.~A.}\ \bibnamefont {Anchordoqui}}, \bibinfo {author} {\bibfnamefont {I.}~\bibnamefont {Antoniadis}}, \bibinfo {author} {\bibfnamefont {D.}~\bibnamefont {Bielli}}, \bibinfo {author} {\bibfnamefont {A.}~\bibnamefont {Chatrabhuti}}, \ and\ \bibinfo {author} {\bibfnamefont {H.}~\bibnamefont {Isono}},\ }\href@noop {} {\  (\bibinfo {year} {2024}{\natexlab{c}})},\ \Eprint {http://arxiv.org/abs/2410.18649} {arXiv:2410.18649 [hep-th]} \BibitemShut {NoStop}%
\bibitem [{\citenamefont {Awad}\ \emph {et~al.}(2018)\citenamefont {Awad}, \citenamefont {El~Hanafy}, \citenamefont {Nashed},\ and\ \citenamefont {Saridakis}}]{Awad:2017yod}%
  \BibitemOpen
  \bibfield  {author} {\bibinfo {author} {\bibfnamefont {A.}~\bibnamefont {Awad}}, \bibinfo {author} {\bibfnamefont {W.}~\bibnamefont {El~Hanafy}}, \bibinfo {author} {\bibfnamefont {G.~G.~L.}\ \bibnamefont {Nashed}}, \ and\ \bibinfo {author} {\bibfnamefont {E.~N.}\ \bibnamefont {Saridakis}},\ }\href {\doibase 10.1088/1475-7516/2018/02/052} {\bibfield  {journal} {\bibinfo  {journal} {JCAP}\ }\textbf {\bibinfo {volume} {02}},\ \bibinfo {pages} {052} (\bibinfo {year} {2018})},\ \Eprint {http://arxiv.org/abs/1710.10194} {arXiv:1710.10194 [gr-qc]} \BibitemShut {NoStop}%
\bibitem [{\citenamefont {Hashim}\ \emph {et~al.}(2021{\natexlab{a}})\citenamefont {Hashim}, \citenamefont {El~Hanafy}, \citenamefont {Golovnev},\ and\ \citenamefont {El-Zant}}]{Hashim:2020sez}%
  \BibitemOpen
  \bibfield  {author} {\bibinfo {author} {\bibfnamefont {M.}~\bibnamefont {Hashim}}, \bibinfo {author} {\bibfnamefont {W.}~\bibnamefont {El~Hanafy}}, \bibinfo {author} {\bibfnamefont {A.}~\bibnamefont {Golovnev}}, \ and\ \bibinfo {author} {\bibfnamefont {A.~A.}\ \bibnamefont {El-Zant}},\ }\href {\doibase 10.1088/1475-7516/2021/07/052} {\bibfield  {journal} {\bibinfo  {journal} {JCAP}\ }\textbf {\bibinfo {volume} {07}},\ \bibinfo {pages} {052} (\bibinfo {year} {2021}{\natexlab{a}})},\ \Eprint {http://arxiv.org/abs/2010.14964} {arXiv:2010.14964 [astro-ph.CO]} \BibitemShut {NoStop}%
\bibitem [{\citenamefont {Hashim}\ \emph {et~al.}(2021{\natexlab{b}})\citenamefont {Hashim}, \citenamefont {El-Zant}, \citenamefont {El~Hanafy},\ and\ \citenamefont {Golovnev}}]{Hashim:2021pkq}%
  \BibitemOpen
  \bibfield  {author} {\bibinfo {author} {\bibfnamefont {M.}~\bibnamefont {Hashim}}, \bibinfo {author} {\bibfnamefont {A.~A.}\ \bibnamefont {El-Zant}}, \bibinfo {author} {\bibfnamefont {W.}~\bibnamefont {El~Hanafy}}, \ and\ \bibinfo {author} {\bibfnamefont {A.}~\bibnamefont {Golovnev}},\ }\href {\doibase 10.1088/1475-7516/2021/07/053} {\bibfield  {journal} {\bibinfo  {journal} {JCAP}\ }\textbf {\bibinfo {volume} {07}},\ \bibinfo {pages} {053} (\bibinfo {year} {2021}{\natexlab{b}})},\ \Eprint {http://arxiv.org/abs/2104.08311} {arXiv:2104.08311 [astro-ph.CO]} \BibitemShut {NoStop}%
\bibitem [{\citenamefont {Yadav}\ \emph {et~al.}(2024)\citenamefont {Yadav}, \citenamefont {Kumar}, \citenamefont {Kibris},\ and\ \citenamefont {Akarsu}}]{Yadav:2024duq}%
  \BibitemOpen
  \bibfield  {author} {\bibinfo {author} {\bibfnamefont {A.}~\bibnamefont {Yadav}}, \bibinfo {author} {\bibfnamefont {S.}~\bibnamefont {Kumar}}, \bibinfo {author} {\bibfnamefont {C.}~\bibnamefont {Kibris}}, \ and\ \bibinfo {author} {\bibfnamefont {O.}~\bibnamefont {Akarsu}},\ }\href@noop {} {\  (\bibinfo {year} {2024})},\ \Eprint {http://arxiv.org/abs/2406.18496} {arXiv:2406.18496 [astro-ph.CO]} \BibitemShut {NoStop}%
\bibitem [{\citenamefont {Nesseris}\ \emph {et~al.}(2013)\citenamefont {Nesseris}, \citenamefont {Basilakos}, \citenamefont {Saridakis},\ and\ \citenamefont {Perivolaropoulos}}]{Nesseris_2013}%
  \BibitemOpen
  \bibfield  {author} {\bibinfo {author} {\bibfnamefont {S.}~\bibnamefont {Nesseris}}, \bibinfo {author} {\bibfnamefont {S.}~\bibnamefont {Basilakos}}, \bibinfo {author} {\bibfnamefont {E.~N.}\ \bibnamefont {Saridakis}}, \ and\ \bibinfo {author} {\bibfnamefont {L.}~\bibnamefont {Perivolaropoulos}},\ }\href {\doibase 10.1103/physrevd.88.103010} {\bibfield  {journal} {\bibinfo  {journal} {Physical Review D}\ }\textbf {\bibinfo {volume} {88}} (\bibinfo {year} {2013}),\ 10.1103/physrevd.88.103010}\BibitemShut {NoStop}%
\bibitem [{\citenamefont {Golovnev}\ and\ \citenamefont {Koivisto}(2018)}]{Golovnev_2018}%
  \BibitemOpen
  \bibfield  {author} {\bibinfo {author} {\bibfnamefont {A.}~\bibnamefont {Golovnev}}\ and\ \bibinfo {author} {\bibfnamefont {T.}~\bibnamefont {Koivisto}},\ }\href {\doibase 10.1088/1475-7516/2018/11/012} {\bibfield  {journal} {\bibinfo  {journal} {Journal of Cosmology and Astroparticle Physics}\ }\textbf {\bibinfo {volume} {2018}},\ \bibinfo {pages} {012–012} (\bibinfo {year} {2018})}\BibitemShut {NoStop}%
\bibitem [{\citenamefont {Sagredo}\ \emph {et~al.}(2018)\citenamefont {Sagredo}, \citenamefont {Nesseris},\ and\ \citenamefont {Sapone}}]{Sagredo_2018}%
  \BibitemOpen
  \bibfield  {author} {\bibinfo {author} {\bibfnamefont {B.}~\bibnamefont {Sagredo}}, \bibinfo {author} {\bibfnamefont {S.}~\bibnamefont {Nesseris}}, \ and\ \bibinfo {author} {\bibfnamefont {D.}~\bibnamefont {Sapone}},\ }\href {\doibase 10.1103/physrevd.98.083543} {\bibfield  {journal} {\bibinfo  {journal} {Physical Review D}\ }\textbf {\bibinfo {volume} {98}} (\bibinfo {year} {2018}),\ 10.1103/physrevd.98.083543}\BibitemShut {NoStop}%
\bibitem [{\citenamefont {Jimenez}\ and\ \citenamefont {Loeb}(2002)}]{Jimenez:2001gg}%
  \BibitemOpen
  \bibfield  {author} {\bibinfo {author} {\bibfnamefont {R.}~\bibnamefont {Jimenez}}\ and\ \bibinfo {author} {\bibfnamefont {A.}~\bibnamefont {Loeb}},\ }\href {\doibase 10.1086/340549} {\bibfield  {journal} {\bibinfo  {journal} {Astrophys. J.}\ }\textbf {\bibinfo {volume} {573}},\ \bibinfo {pages} {37} (\bibinfo {year} {2002})},\ \Eprint {http://arxiv.org/abs/astro-ph/0106145} {arXiv:astro-ph/0106145} \BibitemShut {NoStop}%
\bibitem [{\citenamefont {Moresco}\ \emph {et~al.}(2012)\citenamefont {Moresco}, \citenamefont {Verde}, \citenamefont {Pozzetti}, \citenamefont {Jimenez},\ and\ \citenamefont {Cimatti}}]{Moresco:2012by}%
  \BibitemOpen
  \bibfield  {author} {\bibinfo {author} {\bibfnamefont {M.}~\bibnamefont {Moresco}}, \bibinfo {author} {\bibfnamefont {L.}~\bibnamefont {Verde}}, \bibinfo {author} {\bibfnamefont {L.}~\bibnamefont {Pozzetti}}, \bibinfo {author} {\bibfnamefont {R.}~\bibnamefont {Jimenez}}, \ and\ \bibinfo {author} {\bibfnamefont {A.}~\bibnamefont {Cimatti}},\ }\href {\doibase 10.1088/1475-7516/2012/07/053} {\bibfield  {journal} {\bibinfo  {journal} {JCAP}\ }\textbf {\bibinfo {volume} {07}},\ \bibinfo {pages} {053} (\bibinfo {year} {2012})},\ \Eprint {http://arxiv.org/abs/1201.6658} {arXiv:1201.6658 [astro-ph.CO]} \BibitemShut {NoStop}%
\bibitem [{\citenamefont {Moresco}(2015)}]{Moresco:2015cya}%
  \BibitemOpen
  \bibfield  {author} {\bibinfo {author} {\bibfnamefont {M.}~\bibnamefont {Moresco}},\ }\href {\doibase 10.1093/mnrasl/slv037} {\bibfield  {journal} {\bibinfo  {journal} {Mon. Not. Roy. Astron. Soc.}\ }\textbf {\bibinfo {volume} {450}},\ \bibinfo {pages} {L16} (\bibinfo {year} {2015})},\ \Eprint {http://arxiv.org/abs/1503.01116} {arXiv:1503.01116 [astro-ph.CO]} \BibitemShut {NoStop}%
\bibitem [{\citenamefont {Moresco}\ \emph {et~al.}(2016)\citenamefont {Moresco}, \citenamefont {Pozzetti}, \citenamefont {Cimatti}, \citenamefont {Jimenez}, \citenamefont {Maraston}, \citenamefont {Verde}, \citenamefont {Thomas}, \citenamefont {Citro}, \citenamefont {Tojeiro},\ and\ \citenamefont {Wilkinson}}]{Moresco:2016mzx}%
  \BibitemOpen
  \bibfield  {author} {\bibinfo {author} {\bibfnamefont {M.}~\bibnamefont {Moresco}}, \bibinfo {author} {\bibfnamefont {L.}~\bibnamefont {Pozzetti}}, \bibinfo {author} {\bibfnamefont {A.}~\bibnamefont {Cimatti}}, \bibinfo {author} {\bibfnamefont {R.}~\bibnamefont {Jimenez}}, \bibinfo {author} {\bibfnamefont {C.}~\bibnamefont {Maraston}}, \bibinfo {author} {\bibfnamefont {L.}~\bibnamefont {Verde}}, \bibinfo {author} {\bibfnamefont {D.}~\bibnamefont {Thomas}}, \bibinfo {author} {\bibfnamefont {A.}~\bibnamefont {Citro}}, \bibinfo {author} {\bibfnamefont {R.}~\bibnamefont {Tojeiro}}, \ and\ \bibinfo {author} {\bibfnamefont {D.}~\bibnamefont {Wilkinson}},\ }\href {\doibase 10.1088/1475-7516/2016/05/014} {\bibfield  {journal} {\bibinfo  {journal} {JCAP}\ }\textbf {\bibinfo {volume} {05}},\ \bibinfo {pages} {014} (\bibinfo {year} {2016})},\ \Eprint {http://arxiv.org/abs/1601.01701} {arXiv:1601.01701 [astro-ph.CO]} \BibitemShut {NoStop}%
\bibitem [{\citenamefont {Collaboration}\ \emph {et~al.}(2024{\natexlab{a}})\citenamefont {Collaboration}, \citenamefont {Adame},  \emph {et~al.}}]{desicollaboration2024desi2024iiibaryon}%
  \BibitemOpen
  \bibfield  {author} {\bibinfo {author} {\bibfnamefont {D.}~\bibnamefont {Collaboration}}, \bibinfo {author} {\bibfnamefont {A.~G.}\ \bibnamefont {Adame}}, ,  \emph {et~al.},\ }\href {https://arxiv.org/abs/2404.03000} {\enquote {\bibinfo {title} {Desi 2024 iii: Baryon acoustic oscillations from galaxies and quasars},}\ } (\bibinfo {year} {2024}{\natexlab{a}}),\ \Eprint {http://arxiv.org/abs/2404.03000} {arXiv:2404.03000 [astro-ph.CO]} \BibitemShut {NoStop}%
\bibitem [{\citenamefont {Collaboration}\ \emph {et~al.}(2024{\natexlab{b}})\citenamefont {Collaboration}, \citenamefont {Adame} \emph {et~al.}}]{desicollaboration2024desi2024ivbaryon}%
  \BibitemOpen
  \bibfield  {author} {\bibinfo {author} {\bibfnamefont {D.}~\bibnamefont {Collaboration}}, \bibinfo {author} {\bibfnamefont {A.~G.}\ \bibnamefont {Adame}},  \emph {et~al.},\ }\href {https://arxiv.org/abs/2404.03001} {\enquote {\bibinfo {title} {Desi 2024 iv: Baryon acoustic oscillations from the lyman alpha forest},}\ } (\bibinfo {year} {2024}{\natexlab{b}}),\ \Eprint {http://arxiv.org/abs/2404.03001} {arXiv:2404.03001 [astro-ph.CO]} \BibitemShut {NoStop}%
\bibitem [{\citenamefont {Collaboration}\ \emph {et~al.}(2024{\natexlab{c}})\citenamefont {Collaboration}, \citenamefont {Adame} \emph {et~al.}}]{desicollaboration2024desi2024vicosmological}%
  \BibitemOpen
  \bibfield  {author} {\bibinfo {author} {\bibfnamefont {D.}~\bibnamefont {Collaboration}}, \bibinfo {author} {\bibfnamefont {A.~G.}\ \bibnamefont {Adame}},  \emph {et~al.},\ }\href {https://arxiv.org/abs/2404.03002} {\enquote {\bibinfo {title} {Desi 2024 vi: Cosmological constraints from the measurements of baryon acoustic oscillations},}\ } (\bibinfo {year} {2024}{\natexlab{c}}),\ \Eprint {http://arxiv.org/abs/2404.03002} {arXiv:2404.03002 [astro-ph.CO]} \BibitemShut {NoStop}%
\bibitem [{\citenamefont {Brout}\ \emph {et~al.}(2022)\citenamefont {Brout} \emph {et~al.}}]{Brout:2022vxf}%
  \BibitemOpen
  \bibfield  {author} {\bibinfo {author} {\bibfnamefont {D.}~\bibnamefont {Brout}} \emph {et~al.},\ }\href {\doibase 10.3847/1538-4357/ac8e04} {\bibfield  {journal} {\bibinfo  {journal} {Astrophys. J.}\ }\textbf {\bibinfo {volume} {938}},\ \bibinfo {pages} {110} (\bibinfo {year} {2022})},\ \Eprint {http://arxiv.org/abs/2202.04077} {arXiv:2202.04077 [astro-ph.CO]} \BibitemShut {NoStop}%
\bibitem [{\citenamefont {Collaboration}\ \emph {et~al.}(2024{\natexlab{d}})\citenamefont {Collaboration}, \citenamefont {Abbott}, \citenamefont {Acevedo} \emph {et~al.}}]{descollaboration2024darkenergysurveycosmology}%
  \BibitemOpen
  \bibfield  {author} {\bibinfo {author} {\bibfnamefont {D.}~\bibnamefont {Collaboration}}, \bibinfo {author} {\bibfnamefont {T.~M.~C.}\ \bibnamefont {Abbott}}, \bibinfo {author} {\bibfnamefont {M.}~\bibnamefont {Acevedo}},  \emph {et~al.},\ }\href {https://arxiv.org/abs/2401.02929} {\enquote {\bibinfo {title} {The dark energy survey: Cosmology results with ~1500 new high-redshift type ia supernovae using the full 5-year dataset},}\ } (\bibinfo {year} {2024}{\natexlab{d}}),\ \Eprint {http://arxiv.org/abs/2401.02929} {arXiv:2401.02929 [astro-ph.CO]} \BibitemShut {NoStop}%
\bibitem [{\citenamefont {Aver}\ \emph {et~al.}(2015)\citenamefont {Aver}, \citenamefont {Olive},\ and\ \citenamefont {Skillman}}]{Aver:2015iza}%
  \BibitemOpen
  \bibfield  {author} {\bibinfo {author} {\bibfnamefont {E.}~\bibnamefont {Aver}}, \bibinfo {author} {\bibfnamefont {K.~A.}\ \bibnamefont {Olive}}, \ and\ \bibinfo {author} {\bibfnamefont {E.~D.}\ \bibnamefont {Skillman}},\ }\href {\doibase 10.1088/1475-7516/2015/07/011} {\bibfield  {journal} {\bibinfo  {journal} {JCAP}\ }\textbf {\bibinfo {volume} {07}},\ \bibinfo {pages} {011} (\bibinfo {year} {2015})},\ \Eprint {http://arxiv.org/abs/1503.08146} {arXiv:1503.08146 [astro-ph.CO]} \BibitemShut {NoStop}%
\bibitem [{\citenamefont {Cooke}\ \emph {et~al.}(2018)\citenamefont {Cooke}, \citenamefont {Pettini},\ and\ \citenamefont {Steidel}}]{Cooke:2017cwo}%
  \BibitemOpen
  \bibfield  {author} {\bibinfo {author} {\bibfnamefont {R.~J.}\ \bibnamefont {Cooke}}, \bibinfo {author} {\bibfnamefont {M.}~\bibnamefont {Pettini}}, \ and\ \bibinfo {author} {\bibfnamefont {C.~C.}\ \bibnamefont {Steidel}},\ }\href {\doibase 10.3847/1538-4357/aaab53} {\bibfield  {journal} {\bibinfo  {journal} {Astrophys. J.}\ }\textbf {\bibinfo {volume} {855}},\ \bibinfo {pages} {102} (\bibinfo {year} {2018})},\ \Eprint {http://arxiv.org/abs/1710.11129} {arXiv:1710.11129 [astro-ph.CO]} \BibitemShut {NoStop}%
\bibitem [{\citenamefont {Consiglio}\ \emph {et~al.}(2018)\citenamefont {Consiglio}, \citenamefont {de~Salas}, \citenamefont {Mangano}, \citenamefont {Miele}, \citenamefont {Pastor},\ and\ \citenamefont {Pisanti}}]{Consiglio:2017pot}%
  \BibitemOpen
  \bibfield  {author} {\bibinfo {author} {\bibfnamefont {R.}~\bibnamefont {Consiglio}}, \bibinfo {author} {\bibfnamefont {P.~F.}\ \bibnamefont {de~Salas}}, \bibinfo {author} {\bibfnamefont {G.}~\bibnamefont {Mangano}}, \bibinfo {author} {\bibfnamefont {G.}~\bibnamefont {Miele}}, \bibinfo {author} {\bibfnamefont {S.}~\bibnamefont {Pastor}}, \ and\ \bibinfo {author} {\bibfnamefont {O.}~\bibnamefont {Pisanti}},\ }\href {\doibase 10.1016/j.cpc.2018.06.022} {\bibfield  {journal} {\bibinfo  {journal} {Comput. Phys. Commun.}\ }\textbf {\bibinfo {volume} {233}},\ \bibinfo {pages} {237} (\bibinfo {year} {2018})},\ \Eprint {http://arxiv.org/abs/1712.04378} {arXiv:1712.04378 [astro-ph.CO]} \BibitemShut {NoStop}%
\bibitem [{\citenamefont {Blas}\ \emph {et~al.}(2011)\citenamefont {Blas}, \citenamefont {Lesgourgues},\ and\ \citenamefont {Tram}}]{Diego_Blas_2011}%
  \BibitemOpen
  \bibfield  {author} {\bibinfo {author} {\bibfnamefont {D.}~\bibnamefont {Blas}}, \bibinfo {author} {\bibfnamefont {J.}~\bibnamefont {Lesgourgues}}, \ and\ \bibinfo {author} {\bibfnamefont {T.}~\bibnamefont {Tram}},\ }\href {\doibase 10.1088/1475-7516/2011/07/034} {\bibfield  {journal} {\bibinfo  {journal} {Journal of Cosmology and Astroparticle Physics}\ }\textbf {\bibinfo {volume} {2011}},\ \bibinfo {pages} {034–034} (\bibinfo {year} {2011})}\BibitemShut {NoStop}%
\bibitem [{\citenamefont {Brinckmann}\ and\ \citenamefont {Lesgourgues}(2018)}]{brinckmann2018montepython3boostedmcmc}%
  \BibitemOpen
  \bibfield  {author} {\bibinfo {author} {\bibfnamefont {T.}~\bibnamefont {Brinckmann}}\ and\ \bibinfo {author} {\bibfnamefont {J.}~\bibnamefont {Lesgourgues}},\ }\href {https://arxiv.org/abs/1804.07261} {\enquote {\bibinfo {title} {Montepython 3: boosted mcmc sampler and other features},}\ } (\bibinfo {year} {2018}),\ \Eprint {http://arxiv.org/abs/1804.07261} {arXiv:1804.07261 [astro-ph.CO]} \BibitemShut {NoStop}%
\bibitem [{\citenamefont {Audren}\ \emph {et~al.}(2013)\citenamefont {Audren}, \citenamefont {Lesgourgues}, \citenamefont {Benabed},\ and\ \citenamefont {Prunet}}]{Benjamin_Audren_2013}%
  \BibitemOpen
  \bibfield  {author} {\bibinfo {author} {\bibfnamefont {B.}~\bibnamefont {Audren}}, \bibinfo {author} {\bibfnamefont {J.}~\bibnamefont {Lesgourgues}}, \bibinfo {author} {\bibfnamefont {K.}~\bibnamefont {Benabed}}, \ and\ \bibinfo {author} {\bibfnamefont {S.}~\bibnamefont {Prunet}},\ }\href {\doibase 10.1088/1475-7516/2013/02/001} {\bibfield  {journal} {\bibinfo  {journal} {Journal of Cosmology and Astroparticle Physics}\ }\textbf {\bibinfo {volume} {2013}},\ \bibinfo {pages} {001–001} (\bibinfo {year} {2013})}\BibitemShut {NoStop}%
\bibitem [{\citenamefont {Murakami}\ \emph {et~al.}(2023)\citenamefont {Murakami}, \citenamefont {Riess}, \citenamefont {Stahl}, \citenamefont {Kenworthy}, \citenamefont {Pluck}, \citenamefont {Macoretta}, \citenamefont {Brout}, \citenamefont {Jones}, \citenamefont {Scolnic},\ and\ \citenamefont {Filippenko}}]{Murakami:2023xuy}%
  \BibitemOpen
  \bibfield  {author} {\bibinfo {author} {\bibfnamefont {Y.~S.}\ \bibnamefont {Murakami}}, \bibinfo {author} {\bibfnamefont {A.~G.}\ \bibnamefont {Riess}}, \bibinfo {author} {\bibfnamefont {B.~E.}\ \bibnamefont {Stahl}}, \bibinfo {author} {\bibfnamefont {W.~D.}\ \bibnamefont {Kenworthy}}, \bibinfo {author} {\bibfnamefont {D.-M.~A.}\ \bibnamefont {Pluck}}, \bibinfo {author} {\bibfnamefont {A.}~\bibnamefont {Macoretta}}, \bibinfo {author} {\bibfnamefont {D.}~\bibnamefont {Brout}}, \bibinfo {author} {\bibfnamefont {D.~O.}\ \bibnamefont {Jones}}, \bibinfo {author} {\bibfnamefont {D.~M.}\ \bibnamefont {Scolnic}}, \ and\ \bibinfo {author} {\bibfnamefont {A.~V.}\ \bibnamefont {Filippenko}},\ }\href {\doibase 10.1088/1475-7516/2023/11/046} {\bibfield  {journal} {\bibinfo  {journal} {JCAP}\ }\textbf {\bibinfo {volume} {11}},\ \bibinfo {pages} {046} (\bibinfo {year} {2023})},\ \Eprint {http://arxiv.org/abs/2306.00070} {arXiv:2306.00070 [astro-ph.CO]} \BibitemShut {NoStop}%
\end{thebibliography}%

\end{document}